\documentclass[12pt,nofootinbib]{article}
\pdfoutput=1
\usepackage{slashed}
\usepackage{amsmath,amssymb,graphicx,multicol} 
\usepackage{epsf}
\usepackage[nosort]{cite}
\usepackage{cancel}
\usepackage{framed}
\usepackage{multirow}
\usepackage{color}
\usepackage{subfigure}
\usepackage{tikz}
\usepackage{hyperref}
\usepackage{fixltx2e}

\usetikzlibrary{arrows,shapes} \usetikzlibrary{trees} \usetikzlibrary{decorations.pathmorphing}
\usetikzlibrary{decorations.markings} \usetikzlibrary{calc}

\newcommand{\hhref}[1]{\href{http://arxiv.org/abs/#1}{arXiv:#1}}

\newcommand{\beq}{\begin{eqnarray}}
\newcommand{\eeq}{\end{eqnarray}}

\newcommand{\centeron}[2]{{\setbox0=\hbox{#1}\setbox1=\hbox{#2}\ifdim

\wd1>\wd0\kern.5\wd1\kern-.5\wd0\fi \copy0

\kern-.5\wd0\kern-.5\wd1\copy1\ifdim\wd0>\wd1
                                       \kern.5\wd0\kern-.5\wd1\fi}}
\newcommand{\ltap}{\>\centeron{\raise.35ex\hbox{$<$}}
                               {\lower.65ex\hbox{$\sim$}}\>}
\newcommand{\gtap}{\>\centeron{\raise.35ex\hbox{$>$}}
                               {\lower.65ex\hbox{$\sim$}}\>}

\newcommand\ZZ{\hbox{\zfont Z\kern-.4emZ}}
\font\zfont = cmss10 

\newcommand{\tn}[1]{\ensuremath{\textnormal{#1}}}

\begin{document}
\tikzset{
  vector/.style={decorate, decoration={snake}, draw}, %
  provector/.style={decorate, decoration={snake,amplitude=2.5pt}, draw},%
  antivector/.style={decorate, decoration={snake,amplitude=-2.5pt}, draw},%
  fermion/.style={draw=black, postaction={decorate}, decoration={markings,mark=at position .55 with {\arrow[draw=black]{>}}}},%
  fermionbar/.style={draw=black, postaction={decorate}, decoration={markings,mark=at position .55
      with {\arrow[draw=black]{<}}}}, %
  fermionnoarrow/.style={draw=black}, %
  gluon/.style={decorate, draw=black, decoration={coil,amplitude=4pt, segment length=5pt}},%
  scalar/.style={dashed,draw=black, postaction={decorate}, },%
  scalarbar/.style={dashed,draw=black, postaction={decorate}, decoration={markings,mark=at position .55 with {\arrow[draw=black]{<}}}},%
  scalarnoarrow/.style={dashed,draw=black}, %
  bigvector/.style={decorate, decoration={snake,amplitude=4pt}, draw},%
 }

\begin{titlepage}
\vspace*{-2.5cm}
\begin{flushright}
{\small
CERN-PH-TH/2013-292\\
DESY 13-233
}
\end{flushright}
\vspace*{0.5cm}

\begin{center}
{\Large \bf   Very boosted Higgs in gluon fusion}
\end{center}
\vskip0.5cm

\renewcommand{\thefootnote}{\fnsymbol{footnote}}

\begin{center}
{\large C.~Grojean$^{\, a}$, E.~Salvioni$^{\, b,c,d}$, M.~Schlaffer$^{\, e}$ and A.~Weiler$^{\, c,e\,\,}$\footnote{Emails: christophe.grojean@cern.ch, esalvioni@ucdavis.edu, matthias.schlaffer@desy.de,\\ andreas.weiler@cern.ch}}
\end{center}

\vskip 10pt

\begin{center}
\centerline{$^{a}$ {\small \it ICREA at IFAE, Universitat Aut\'onoma de Barcelona, E-08193 Bellaterra, Spain}}
\vskip 5pt
\centerline{$^{b}${\small \it Department of Physics, University of California, Davis, CA 95616, USA}}
\vskip 5pt
\centerline{$^{c}${\small \it Theory Division, Physics Department, CERN, CH-1211 Geneva 23, Switzerland}}
\vskip 5pt
\centerline{$^{d}${\small \it Dipartimento di Fisica e Astronomia, Universit\`a di Padova and}}
\centerline{{\small \it INFN, Sezione di Padova, Via Marzolo 8, I-35131 Padova, Italy}}
\vskip 5pt
\centerline{$^{e}${\small \it DESY, Notkestrasse 85, D-22607 Hamburg, Germany}}
\end{center}

\begin{abstract}
\noindent 
The Higgs production and decay rates offer a new way to probe new physics beyond the Standard Model. While dynamics aiming at alleviating the hierarchy problem generically predict deviations in the Higgs rates, the current experimental analyses cannot resolve the long- and short-distance contributions to the gluon fusion process and thus cannot access directly the coupling between the Higgs and the top quark. We investigate the production of a boosted Higgs in association with a high-transverse momentum jet as an alternative to the $t\bar t h$ channel to pin down this crucial coupling. Presented first in the context of an effective field theory, our analysis is then applied to models of partial compositeness at the TeV scale and of natural supersymmetry.
\end{abstract}

\end{titlepage}

\section{Introduction}

With the discovery~\cite{ATLASdiscovery, CMSdiscovery} last year of a new resonance, whose properties so far show no deviations from those of the long sought-after Higgs boson, a new era started in the understanding of the Standard Model (SM) of particle physics. In the absence of any evidence for any other new degree of freedom at the weak scale, a mass gap is likely to separate the SM particles from the dynamics generating and stabilizing the Higgs potential. Our ignorance about the new physics sector can thus be conveniently parametrized in terms of a set of higher dimensional operators built out of SM fields obeying the SM symmetries. For a single family of fermions, there are 59 independent ways to deform the SM~\cite{dim6}. Of particular interest are the 30 deformation directions affecting Higgs physics~\cite{Higgsdim6, Elias-Miro:2013mua}. Actually, $20$ of them were already constrained at the per-mil/per cent level before the Higgs discovery itself, thanks to electroweak measurements involving massive gauge bosons and bounds on quark and lepton dipole moments. The Higgs data collected by ATLAS and CMS  (as well as by the Tevatron experiments) start putting interesting bounds on the remaining 8 $CP$-even and 2 $CP$-odd directions~\cite{Elias-Miro:2013mua,Pomarol:2013zra}. In this regard, a Higgs mass around 125\,GeV offers remarkable opportunities to probe these directions, since it opens numerous decay channels with a rate accessible with the current integrated luminosity delivered by the LHC. 

Among these operators, four are particularly important, since they control the main production mechanism of the Higgs, namely gluon fusion. These are\footnote{$G^a_{\mu\nu}$ denotes the QCD gauge-field strength and $\widetilde{G}^{a\,\mu\nu}=(1/2)\epsilon^{\mu\nu\rho\sigma}G_{\rho\sigma}^{a}$ its dual. $\alpha_s$ is the QCD coupling strength. $H$ is the SM Higgs doublet and $\tilde H \equiv i \sigma_2 H^*$, $v$ is the SM Higgs vacuum expectation value related to the Fermi constant by $v=( \sqrt{2}\,G_F )^{-1/2} \simeq 246\,\textrm{GeV}$, $y_t$ is the SM top Yukawa coupling, and finally $Q_L$ and $t_R$ are the $SU(2)_L$ quark doublet and charge-2/3 quark singlet of the third generation.}
\begin{align} \nonumber
\label{eq:operatorOg}
\mathcal{O}_y \,=&\, \frac{y_t}{v^2} |H|^2 {\bar Q}_L \tilde H t_R\,\, ,
\qquad \quad\quad
\mathcal{O}_H = \frac{1}{2v^2} \partial_{\mu}|H|^2 \partial^{\mu}|H|^2\,, \\ 
\mathcal{O}_g \,=&\, \frac{\alpha_s}{12 \pi v^2} |H|^2 G_{\mu\nu}^a G^{a\, \mu\nu}\,,\qquad \, \widetilde{\mathcal{O}}_g = \frac{\alpha_s}{8 \pi v^2} |H|^2 G_{\mu\nu}^a \widetilde{G}^{a\, \mu\nu}
\,,
\end{align}
added to the SM as
\begin{equation} \label{dim6}
\mathcal{L}_{\mathrm{eff}}=\mathcal{L}_{\mathrm{SM}}+\left(c_{y}\mathcal{O}_{y}+\mathrm{h.c.}\right)+c_{H}\mathcal{O}_{H}+c_{g}\mathcal{O}_{g}+\tilde{c}_{g}\widetilde{\mathcal{O}}_{g}\,.
\end{equation}
From this Lagrangian we can extract the terms relevant to Higgs production in gluon fusion,\footnote{See Ref.~\cite{Harlander:2013oja} for a study of the effects of Higgs-gluons interactions of dimension $7$ in $h+\mathrm{jet}$ and $h+2\,\mathrm{jets}$ observables.}
\begin{equation} \label{lagrangian}
-\kappa_t \, \frac{m_{t}}{v} \,  \bar t th +\kappa_{g}\frac{\alpha_{s}}{12\pi}\frac{h}{v}G_{\mu\nu}^{a}G^{\mu\nu\,a}+i\tilde{\kappa}_{t}\frac{m_{t}}{v}\overline{t}\gamma_{5}t h + \tilde{\kappa}_{g}\frac{\alpha_{s}}{8\pi}\frac{h}{v}G_{\mu\nu}^{a}\widetilde{G}^{a\,\mu\nu}+ \mathcal{L}_{\mathrm{QCD}}
\,, 
\end{equation}
where at linear order in the coefficients multiplying the operators in Eq.~\eqref{dim6} we find 
\begin{equation}
\kappa_{t}=1-\mathrm{Re}(c_{y})-\frac{c_{H}}{2}\,,\qquad \kappa_{g}=c_{g}\,,\qquad \tilde{\kappa}_{t}=\mathrm{Im}(c_{y})\,,\qquad \tilde{\kappa}_{g}=\tilde{c}_{g}\,.
\end{equation}
The first two terms in Eq.~\eqref{lagrangian} are invariant under $CP$, whereas the third and fourth terms violate $CP$. We will be mainly interested in $CP$-conserving effects, therefore we set $\tilde{\kappa}_{t}=\tilde{\kappa}_{g}=0$ in the following. We will however return briefly to the $CP$-odd case later (notice that since there is no interference between the $CP$-even and -odd amplitudes for $pp\to h + \mathrm{jet}$, the two cases can be discussed separately). The coefficient $\kappa_{t}$ corrects the top Yukawa coupling, which controls the top loop contribution to $gg\to h$ production, whereas the coefficient $\kappa_{g}$ gives a direct contribution to this process. Unfortunately, the lightness of the Higgs boson plays a malicious role and makes it impossible to disentangle short- and long-distance contribution to the gluon fusion total rate. This limitation is embodied in the Higgs low energy theorem~\cite{HLET} that ensures that the inclusive production cross section is proportional to $(\kappa_{t}+\kappa_{g})^{2}\,$. In fact, from Eq.~\eqref{lagrangian} we readily obtain
\begin{equation}
\frac{\sigma_{\mathrm{incl}}(\kappa_{t},\kappa_{g})}{\sigma_{\mathrm{incl}}^{\mathrm{SM}}}\simeq (\kappa_t + \kappa_g)^2 \left(1 - \frac{7}{15}\,\frac{\kappa_{g}}{\kappa_{t}+\kappa_{g}}\,\frac{m_{h}^{2}}{4m_{t}^{2}}\right) \simeq (\kappa_t+\kappa_g)^2\,,
\end{equation}
where in the second equality we neglected corrections that will remain untouchable at the LHC~\cite{Gillioz:2012se}. The only hope then to have a direct access to the top Yukawa coupling is through the $pp\to t\bar{t} h$ process, which is notoriously difficult due to its high threshold,  its small rate and its complicated final state with copious decay products. Exploratory studies concluded that the sensitivity in the measurement of the top Yukawa coupling through the $t\bar{t}h$ process will be limited to about 10\% within the LHC high-luminosity program~\cite{HL-LHC} and Higgs factories like a 500-GeV ILC or a 80/100-km TLEP will be needed to bring this sensitivity down to a few per-cent level~\cite{ILC-TLEP}. 

In this paper we propose an alternative way to break the degeneracy between the two coefficients $\kappa_{t}$ and $\kappa_{g}$ and thus to provide direct information on the top Yukawa coupling. Our idea relies on the use of extra radiation in the $gg\to h$ process that will allow us to explore the structure of the top loop. When the extra radiation carries away a large amount of energy and boosts the Higgs boson, the process effectively probes the ultraviolet structure of the top loop. Notice that the extra radiation cannot be in the form of a photon, as the amplitude for $gg\to h + \gamma$ vanishes due to Furry's theorem. One is therefore led to consider the production of $h$ in association with a jet.\footnote{The process $gg\to hZ$ is also interesting, despite being subleading with respect to the tree-level contribution from $q\bar{q}\to hZ\,$. See Refs.~\cite{Harlander:2013mla,Englert:2013vua} for very recent studies.} 

Resolving the degeneracy between the two contributions to gluon-fusion Higgs production is not a mere academic
exercise. There exist some new physics scenarios whose identification is endangered by the limitation of the LHC to separately measure $\kappa_t$ and $\kappa_g$. The prime examples are composite Higgs models: the top sector is enlarged and encompasses extra heavy vector-like fermions that mix with the SM top quark. These top partners have a double effect to simultaneously shift the top Yuakwa coupling and to also give rise to a contact interaction between the gluons and the Higgs boson. As it was first noticed in Ref.~\cite{Falkowski:2007hz} and later explored in Refs.~\cite{Low:2010mr, Azatov:2011qy, Delaunay:2013iia, Montull:2013mla}, these two effects precisely annihilate in the $gg\to h$ process in minimal (and most popular) models. Another cancellation can occur in natural supersymmetric (SUSY) scenarios, when a large mixing between the stops conspires to nullify the total stop contribution to $gg\to h$. Even though this is a particular region in the SUSY parameter space, an accumulation of experimental hints might point to such a situation and some experimental ingenuity will be required to directly see the low-lying stops~\cite{Delgado:2012eu}. Thus new physics might well be present in the data already collected at the LHC, while the measurements performed up to now remain blind to it. We advocate that the $pp\to h+\textrm{jet}$ channel can be competitive or at least complementary with the more studied $pp\to t\bar{t}h$ channel to unveil hidden new physics. For previous studies addressing the use of the Higgs transverse momentum distribution as a handle on New Physics, see for example Refs.~\cite{Langenegger:2006wu,Arnesen:2008fb,Bagnaschi:2011tu}.

This paper is organized as follows. In Section~\ref{sec:analysis} we present our analysis of the $pp \to h+ \textrm{jet}$ process, by adopting an effective theory approach to describe the effects of heavy new physics. In order to obtain an estimate of the LHC potential to disentangle the current degeneracy in the plane of effective couplings, we focus on the promising channel where the Higgs boson decays to two collimated tau leptons. Section~\ref{sec:MCHM} studies the ability of the boosted Higgs regime to probe the spectrum of top partners in composite Higgs models, whereas Section~\ref{sec:SUSY} looks at the $ h+ \textrm{jet}$ process as a way to probe light stops in supersymmetric extensions of the SM. Finally, Section~\ref{sec:conclusions} collects our conclusions. We also include an Appendix, where formulae for the $pp\to h+\mathrm{jet}$ cross section mediated by $CP$-violating couplings are reported.

\section{Analysis of $pp \to h + {\rm jet}$}
\label{sec:analysis}
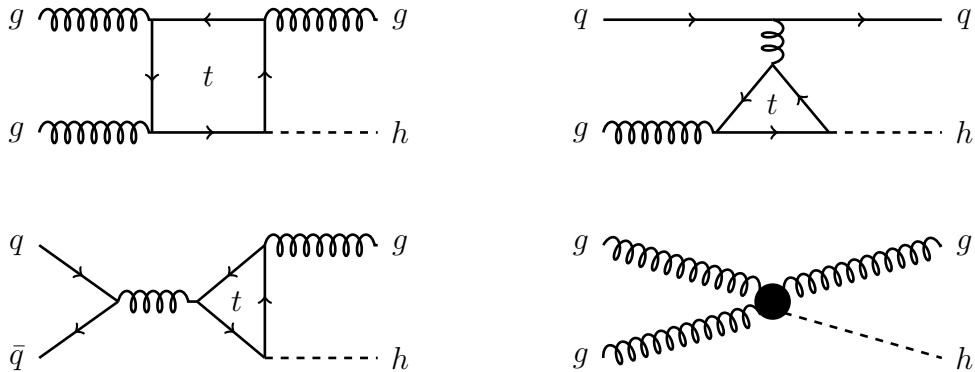
\begin{figure}[thb]
  \centering
  \begin{tikzpicture}[baseline=(current bounding box.center), line width=1 pt, scale=1.5]

    \begin{scope}[shift={(-2.5,0)}]
      \draw[gluon] (-1.5,0.5)--(-0.5,0.5); %
      \node at (-1.7,0.5) {\(g\)}; %
      \draw[gluon] (-1.5,-0.5)--(-0.5,-0.5); %
      \node at (-1.7,-0.5) {\(g\)}; %
      \draw[gluon] (0.5,0.5)--(1.5,0.5); %
      \node at (1.7,0.5) {\(g\)}; %
      \draw[scalar] (0.5,-0.5)--(1.5,-0.5); %
      \node at (1.7,-0.5) {\(h\)}; %
      
      \draw[fermion] (-0.5,0.5)--(-0.5,-0.5); %
      \draw[fermion] (-0.5,-0.5)--(0.5,-0.5); %
      \draw[fermion] (0.5,-0.5)--(0.5,0.5); %
      \draw[fermion] (0.5,0.5)--(-0.5,0.5); %
      \node at (0,0) {\(t\)}; %
    \end{scope}

    \begin{scope}[shift={(2.5,0)}]
      \draw[fermion] (-1.5,0.5)--(0.0,0.5); %
      \node at (-1.7,0.5) {\(q\)}; %
      \draw[fermion] (0.0,0.5)--(1.5,0.5); %
      \node at (1.7,0.5) {\(q\)}; %
      \draw[gluon] (-1.5,-0.5)--(-0.5,-0.5); %
      \node at (-1.7,-0.5) {\(g\)}; %
      \draw[scalar] (0.5,-0.5)--(1.5,-0.5); %
      \node at (1.7,-0.5) {\(h\)}; %

      \draw[gluon] (0.0,0.5)--(0.0,0.1); %
      \draw[fermion] (-0.5,-0.5)--(0.5,-0.5); %
      \draw[fermion] (0.5,-0.5)--(0.0,0.1); %
      \draw[fermion] (0.0,0.1)--(-0.5,-0.5); %
      \node at (0.0,-0.25) {\(t\)}; %
    \end{scope}

    \begin{scope}[shift={(-2.5,-2)}]
      \draw[fermion] (-1.5,0.5)--(-0.8,0.0); %
      \node at (-1.7,0.5) {\(q\)}; %
      \draw[fermion] (-0.8,0.0)--(-1.5,-0.5); %
      \node at (-1.7,-0.5) {\(\bar{q}\)}; %
      \draw[gluon] (0.5,0.5)--(1.5,0.5); %
      \node at (1.7,0.5) {\(g\)}; %
      \draw[scalar] (0.5,-0.5)--(1.5,-0.5); %
      \node at (1.7,-0.5) {\(h\)}; %

      \draw[gluon] (-0.8,0.0)--(-0.1,0.0); %
      \draw[fermion] (-0.1,0.0)--(0.5,-0.5); %
      \draw[fermion] (0.5,-0.5)--(0.5,0.5); %
      \draw[fermion] (0.5,0.5)--(-0.1,0.0); %
      \node at (0.25,0.0) {\(t\)}; %
    \end{scope}

    \begin{scope}[shift={(2.5,-2)}]
      \draw[gluon] (-1.5,0.5)--(-0.1,0.1);%
      \node at (-1.7,0.5) {\(g\)}; %
      \draw[gluon] (-1.5,-0.5)--(-0.1,-0.1); %
      \node at (-1.7,-0.5) {\(g\)}; %
      \draw[gluon] (0.1,0.1)--(1.5,0.5); %
      \node at (1.7,0.5) {\(g\)}; %
      \draw[scalar] (0.1,-0.1)--(1.5,-0.5); %
      \node at (1.7,-0.5) {\(h\)}; %
      \draw[color=black, fill=black] (0,0) circle (0.15); %
    \end{scope}
  \end{tikzpicture}
  \caption{Example Feynman diagrams for \(pp\to h+\tn{jet}\) in the SM and with the contact term.}
  \label{fig:SMdiag}
\end{figure}

%
At the parton level, three subprocesses contribute to the $pp\to h+\tn{jet}$ cross section: these are $gg,qg,q\bar{q}\to h+\tn{jet}$.\footnote{For brevity, we denote the sum $qg + \bar{q}g$ by $qg$.} The expressions of the SM matrix elements for \(gg\to hg\) and \(q\bar{q}\to hg\), mediated by quark loops, were first calculated at LO in QCD in Ref.~\cite{Ellis:1987xu} and shortly after with a different notation in Ref.~\cite{Baur:1989cm}, which we used for our calculations. The matrix element for the $qg\to hq$ process is obtained from the one of \(q\bar{q}\to hg\) by crossing. Some of the Feynman diagrams contributing to $pp\to h+\tn{jet}$ are shown in Fig.~\ref{fig:SMdiag}. When the Lagrangian in Eq.~\eqref{lagrangian} is considered, the top contribution to the amplitudes is simply given by the SM one rescaled by the modified coupling $\kappa_{t}$.\footnote{In the SM, the effect of including the bottom quark contribution in addition to the dominant one due to the top is only of a few percent, if the cut on the transverse momentum is larger than \(50\, \tn{GeV}\)~\cite{Bagnaschi:2011tu,Mantler:2012bj,Grazzini:2013mca}. Since we are interested in larger Higgs transverse momenta, we consistently neglect the bottom in our calculation.}
On the other hand, the contribution of heavy top partners in the loop is described by the effective interaction parameterized by
$\kappa_{g}$, which generates Feynman diagrams such as the lower-right one in Fig.~\ref{fig:SMdiag}. Roughly speaking, this description is reliable as long as the mass of the heavy states is larger than the transverse momentum cut applied, see Section~\ref{sec:MCHM} for a more precise assessment. The corresponding matrix element is obtained from the SM one by sending to infinity the mass of the quark running in the loop. Thus the matrix element squared for each partonic subprocess can be written as
\begin{equation}
\left|\mathcal{M}\right|^{2}\propto \left|\kappa_{t}\,\mathcal{M}_{\mathrm{IR}}(m_{t}) + \kappa_{g}\,\mathcal{M}_{\mathrm{UV}}\right|^{2},
\end{equation}
where $\mathcal{M}_{\mathrm{IR}}$ denotes the amplitude mediated by top loops, and $\mathcal{M}_{\mathrm{UV}}$ the amplitude mediated by the effective point-like interaction. It follows that the hadronic cross section for $pp\to h+\tn{jet}$ can be written as a quadratic polynomial in $\kappa_{t}$ and $\kappa_{g}\,$. Given a transverse momentum cut $p_{T}^{\mathrm{min}}$ and summing over all partonic subprocesses, we can write 
\begin{equation}
  \label{eq:semi-numerical}
  \frac{\sigma_{p_T^{\mathrm{min}}}(\kappa_t,\kappa_g)}{\sigma^{\mathrm{SM}}_{p_T^{\mathrm{min}}}}= (\kappa_t+\kappa_g)^2+\delta\, \kappa_t\, \kappa_g +
  \epsilon \, \kappa_g^2
\end{equation}
where $\sigma$ is the cross section for $pp\to h+\tn{jet}$ and the numerical coefficients $\{\delta\,,\epsilon\}$ depend on $p_{T}^{\mathrm{min}}$. Their values are listed in Table~\ref{tab:coeff} for an LHC center of mass energy of \(\sqrt{s}=14\,\tn{TeV}\) and various choices of $p_{T}^{\mathrm{min}}$. In the calculation, we set the factorization and renormalization scales to the Higgs transverse mass,
\(\mu_{\mathrm{fact}}=\mu_{\mathrm{ren}}=m_T(h)=\sqrt{m_h^2 + p_T^2}\), where \(m_h\) is the mass of the Higgs and $p_T^2=\hat{t}\hat{u}/\hat{s}\,$ its transverse momentum squared. The calculation of the strong coupling constant and the convolution with the PDFs was done using the MSTW 2008 LO PDFs~\cite{Martin:2009iq}. The values of the scalar integrals were obtained from LoopTools-2.8~\cite{Hahn:1998yk}. We have compared our results for the SM cross sections with MCFM-6.6~\cite{Campbell:MCFM} and with HIGLU~\cite{Spira:1995mt}, finding exact agreement.   
\begin{table}
  \centering
  \begin{tabular}{|c|r|r|r|r|r||r|r|r|}
    \hline
    \(\sqrt{s}\) [TeV]&
    \(p_T^{\mathrm{min}}\) [GeV] &\multicolumn{1}{c|}{\(\sigma^{\mathrm{SM}}_{p_T^{\mathrm{min}}}\,[\mathrm{fb}]\)}&\multicolumn{1}{c|}{\(\delta\)}&\multicolumn{1}{c|}{\(\epsilon\)}&
    \(gg,\, qg\) [\%]&\multicolumn{1}{c|}{\(\tilde{\gamma}\cdot 10^2\)}&\multicolumn{1}{c|}{\(\tilde{\delta}\)}&\multicolumn{1}{c|}{\(\tilde{\epsilon}\)}\\\hline 
    \multirow{15}{*}{14}
    &100& 2180&  0.0031&  0.031& 67,\,31&2.6 & 0.033& 0.031\\
    &150& 837 & 0.070&  0.13& 66,\,32&1.7& 0.094& 0.13\\
    &200& 351&  0.20&  0.30& 65,\,34&0.28& 0.22& 0.30\\
    &250& 157&  0.39&  0.56& 63,\,36&0.20& 0.41& 0.56\\
    &300& 74.9&  0.61&  0.89& 61,\,38&1.0 & 0.64& 0.89\\
    &350& 37.7&  0.85&  1.3& 58,\,41&2.2& 0.91& 1.3\\
    &400& 19.9&  1.1&  1.7& 56,\,43&3.4& 1.2& 1.7\\
    &450& 10.9&  1.4&  2.3& 54,\,45&4.6 & 1.5& 2.3\\
    &500& 6.24&  1.7&  2.9& 52,\,47&5.6& 1.8&2.9\\
    &550& 3.68&  2.0&  3.6& 50,\,49&6.5& 2.2&3.6\\
    &600& 2.22&  2.3&  4.4& 48,\,51&7.3& 2.5& 4.4\\
    &650& 1.38&  2.6&  5.2& 46,\,53&7.9& 2.9& 5.2\\
    &700& 0.871&  3.0&  6.2& 45,\,54&8.4& 3.2& 6.2\\
    &750& 0.562&  3.3&  7.2& 43,\,56&8.8& 3.6& 7.2\\
    &800& 0.368&  3.7&  8.4& 42,\,57&9.1& 3.9& 8.4\\\hline
    \multirow{2}{*}{100}& 500& 964& 1.8& 3.1& 72,\,28&5.0& 1.9 &3.1\\
    & 2000& 1.01& 14& 78& 56,\,43&7.0&15 &78\\\hline
  \end{tabular}
  \caption{Summary table of the cross sections for $pp\to h+\tn{jet}$ at proton-proton colliders with \(\sqrt{s}=14\,\tn{TeV}\) and \(\sqrt{s}=100\,\tn{TeV}\). The third, fourth and fifth column show, for the given cut on $p_{T} > p_{T}^{\mathrm{min}}$, the parameters of the semi-numerical formula in Eq.~\eqref{eq:semi-numerical}. The sixth column shows the fraction of the SM cross section coming from the partonic subprocesses $gg$ and $qg$ (the contribution of the $q\bar{q}$ channel is always smaller than $2\%$). The last three columns show the values of the parameters of the semi-numerical formula for the $CP$-odd cross sections, Eq.~\eqref{eq:cp odd}.}
  \label{tab:coeff}
\end{table}
In Table~\ref{tab:coeff} we also show the values of $\{\delta\,,\epsilon\}$ for a 100\,TeV proton-proton collider, for two choices of $p_T^{\mathrm{min}}$, to illustrate the possibilities of such a machine. However, since the efficiencies achievable at the very-high-energy collider are not known yet, we do not consider these values further in our analysis. We note that the $gg$ and $qg$ initial states contribute to the total cross section roughly at the same level, with the relative contribution of $qg$ increasing at higher $p_{T}\,$. The $q\bar{q}$ initial state contributes to the total cross section at the level of $1\div 2\%$.

It is interesting to consider also the possibility that the gluon fusion process receives contribution from $CP$-violating new physics. Considering the operators parameterized by $\tilde{\kappa}_{t}$ and $\tilde{\kappa}_{g}\,$ in Eq.~\eqref{lagrangian}, we can write the cross section for $pp\to h+\tn{jet}$ in a form similar to Eq.~\eqref{eq:semi-numerical}
\begin{equation}
  \label{eq:cp odd}
  \frac{\sigma_{p_T^{\mathrm{min}}}(\tilde{\kappa}_t,\tilde{\kappa}_g)}{\sigma^{\mathrm{SM}}_{p_T^{\mathrm{min}}}}=\frac{9}{4}\left[
    (\tilde{\kappa}_t+\tilde{\kappa}_g)^2 +\tilde{\gamma} \, \tilde{\kappa}_t^2+\tilde{\delta}\, \tilde{\kappa}_t\, \tilde{\kappa}_g +
  \tilde{\epsilon} \, \tilde{\kappa}_g^2\right]
\end{equation}
where the numerical coefficients $\{\tilde{\gamma}\,,\tilde{\delta}\,,\tilde{\epsilon}\}$ depend on $p_{T}^{\mathrm{min}}$. Their values are listed in Table~\ref{tab:coeff}, for the same parameter choices we made in the $CP$-conserving case, and were calculated by using the analytical results for the $pp\to h + \mathrm{jet}$ cross section mediated by $CP$-violating couplings, reported in the Appendix. Notice that the overall factor of $9/4$ appearing in Eq.~\eqref{eq:cp odd} is chosen in accordance with the expression of the inclusive cross section, which reads $\sigma_{\mathrm{incl}}(\tilde{\kappa_{t}},\tilde{\kappa}_{g})/\sigma_{\mathrm{incl}}^{\mathrm{SM}}\simeq 9(\tilde{\kappa}_t+\tilde{\kappa}_g)^2/4\,$.

It has been shown~\cite{Kauffman:1993nv} that the effective operators $hG_{\mu\nu}^{a}G^{a\,\mu\nu}$ and $hG_{\mu\nu}^{a}\widetilde{G}^{a\,\mu\nu}$, when added in turn and with the same normalization to the QCD Lagrangian, lead to identical amplitudes squared for the processes $gg,qg,q\bar{q}\to h+\tn{jet}$. This implies that the identity $\epsilon=\tilde{\epsilon}\,$ holds for any value of $p_{T}^{\mathrm{min}}$, as can be seen in Table~\ref{tab:coeff}.
\subsection{Breaking the degeneracy}
We now aim at giving an estimate of the potential of the boosted Higgs measurement to resolve the ambiguity in the plane of the couplings $(\kappa_{t},\kappa_{g})$. While a full analysis would ideally combine several decay modes of the Higgs, a first estimate can be obtained by looking at one channel only. Because, as shown in Table~\ref{tab:coeff}, in order to break the degeneracy we need to consider very large Higgs transverse momenta and therefore small rates, it is natural to consider first the decay channels with the largest branching ratios, namely $h\to b\bar{b},WW,\tau\tau$. Here we focus on the last mode, and we will comment briefly on other possibilities at the end of this section. For a Higgs transverse momentum larger than $500\,\mathrm{GeV}$, the typical angular separation between the two taus is $\Delta R \sim 2m_{h}/p_{T}\lesssim 0.5$. As a consequence, when at least one of the taus decays hadronically, the standard tau-tagging techniques will fail, due to the non-isolation of the hadronic tau candidate(s). However, such `ditau-jets' can be tagged by adapting the usual tau-tagging algorithm, as suggested in Ref.~\cite{Katz:2010iq}, whose efficiencies for signal identification are assumed here.\footnote{Reference~\cite{Katz:2010iq} applied ditau-tagging to the case of a $Z'$ decaying to $Zh$. We make use of the efficiencies reported in their Table~I for a $2\,\mathrm{TeV}$ $Z'$, which gives a Higgs $p_{T}$ roughly similar to the case we are considering. We assume efficiencies that include in addition to the ditau-jet tagging also the reconstruction of the Higgs mass peak, as it seems unavoidable that an experimental analysis would need to exploit that information.} Including the Higgs and tau branching ratios, we obtain the following estimate of the total efficiency        
\begin{equation}
  \label{eq:2}
  \epsilon_{\mathrm{tot}}=\tn{BR}(h\to\tau\tau) \times \left(\sum\limits_{i\,=\,\tau_{\ell}\tau_{\ell},\,\tau_{\ell}\tau_{h},\,\tau_{h}\tau_{h}} \tn{BR}(\tau\tau\to i) \,\epsilon_i\right)\simeq 2\times 10^{-2}
\end{equation}
where we assumed the SM value for $\mathrm{BR}(h\to \tau\tau)$~\cite{Heinemeyer:2013tqa}.

To break the degeneracy in the $(\kappa_{t},\kappa_{g})$ plane that plagues inclusive Higgs production, we need to combine the measurements of both the inclusive and boosted rates. On the one hand, we take the inclusive Higgs production cross section normalized to its SM value
\begin{equation}
  \label{eq:3}
  \mu_{\tn{incl}}(\kappa_t,\kappa_g)
  =\frac{\sigma_{\tn{incl}}(\kappa_t,\kappa_g)}{\sigma_{\tn{incl}}^{\tn{SM}}}
  \simeq \left(\kappa_t + \kappa_g\right)^2\,.
\end{equation}
We assume the large-luminosity LHC scenario with $3\,\mathrm{ab}^{-1}$ of data at 14\,TeV, and therefore we assign to the measurement of $\mu_{\mathrm{incl}}$ a $10\%$ systematic uncertainty and negligible statistical uncertainty. On the other hand, in order to reduce the theory uncertainty, we consider as boosted observable the ratio
\begin{equation}
  \label{eq:4}
\mathcal{R}(\kappa_t,\kappa_g)=\frac{\sigma_{650\,\tn{GeV}}(\kappa_t,\kappa_g)K_{650\,\mathrm{GeV}}}{\sigma_{150\,\tn{GeV}}(\kappa_t,\kappa_g)K_{150\,\mathrm{GeV}}}\,,
\end{equation}
where $K_{p_T^{\mathrm{min}}}$ are the QCD $K$-factors for the SM, computed using MCFM-6.6 (process 204). The transverse momentum cuts of $650$\,GeV and $150$\,GeV were chosen by means of a rough optimization. The ratio $\mathcal{R}$ is stable under scale variations, as can be seen from Table~\ref{tab:scale-variation}. We remark that presently no exact NLO computation of the SM Higgs $p_{T}$ spectrum is available. The known NLO results, implemented in MCFM-6.6, assume the heavy top approximation and cannot therefore be used for our study, where the full dependence on the top mass is crucial. Nevertheless, multiplying the exact LO cross section times the SM $K$-factor computed in the $m_{t}\to\infty$ limit, as in Eq.~\eqref{eq:4}, is the best approximation available at the present time. From this discussion it is clear that an exact NLO computation of the SM Higgs $p_{T}$ spectrum would be very welcome, and we hope that the QCD community will fill this gap in the near future.\footnote{A first step in this direction has been made in Ref.~\cite{Harlander:2012hf}. For recent progress in the predictions for $h+\mathrm{jet}$, see also Refs.~\cite{Boughezal:2013uia,Hoeche:2014lxa}.}  
\begin{table}[htp]
  \centering
  \begin{tabular}[htb]{|c|c|c|c|c|c|}
    \hline
    \(\mu = \mu_{\mathrm{ren}}=\mu_{\mathrm{fac}}\)& \(\sigma_{150\,\tn{GeV}}\) [fb]& \(K_{150\,\tn{GeV}}\)&
    \(\sigma_{650\,\tn{GeV}}\) [fb]& \(K_{650\,\tn{GeV}}\)& \(\mathcal{R}\)\\\hline
    \(m_T/2\) & \(1.2\cdot10^3\) & 1.16 & 2.0 & 1.14 & $1.66\cdot 10^{-3}$\\\hline
    \(m_T\) & \(0.83\cdot10^3\) & 1.41 & 1.4 & 1.44& $1.69\cdot 10^{-3}$\\\hline
    \(2m_T\) & \(0.60\cdot10^3\) & 1.64 & 0.96 & 1.70& $1.66\cdot 10^{-3}$\\\hline
  \end{tabular}
  \caption{Scale dependence of the SM ($\kappa_{t}=1$, $\kappa_{g}=0$) LO cross sections, $K$-factors and the resulting values of the ratio $\mathcal{R}$, defined in Eq.~\eqref{eq:4}.}
  \label{tab:scale-variation}
\end{table}

\begin{figure}[h!]
  \centering \subfigure[\(\mu^{0}_\tn{incl}=0.8\)]{
    \includegraphics[width=0.45\linewidth]{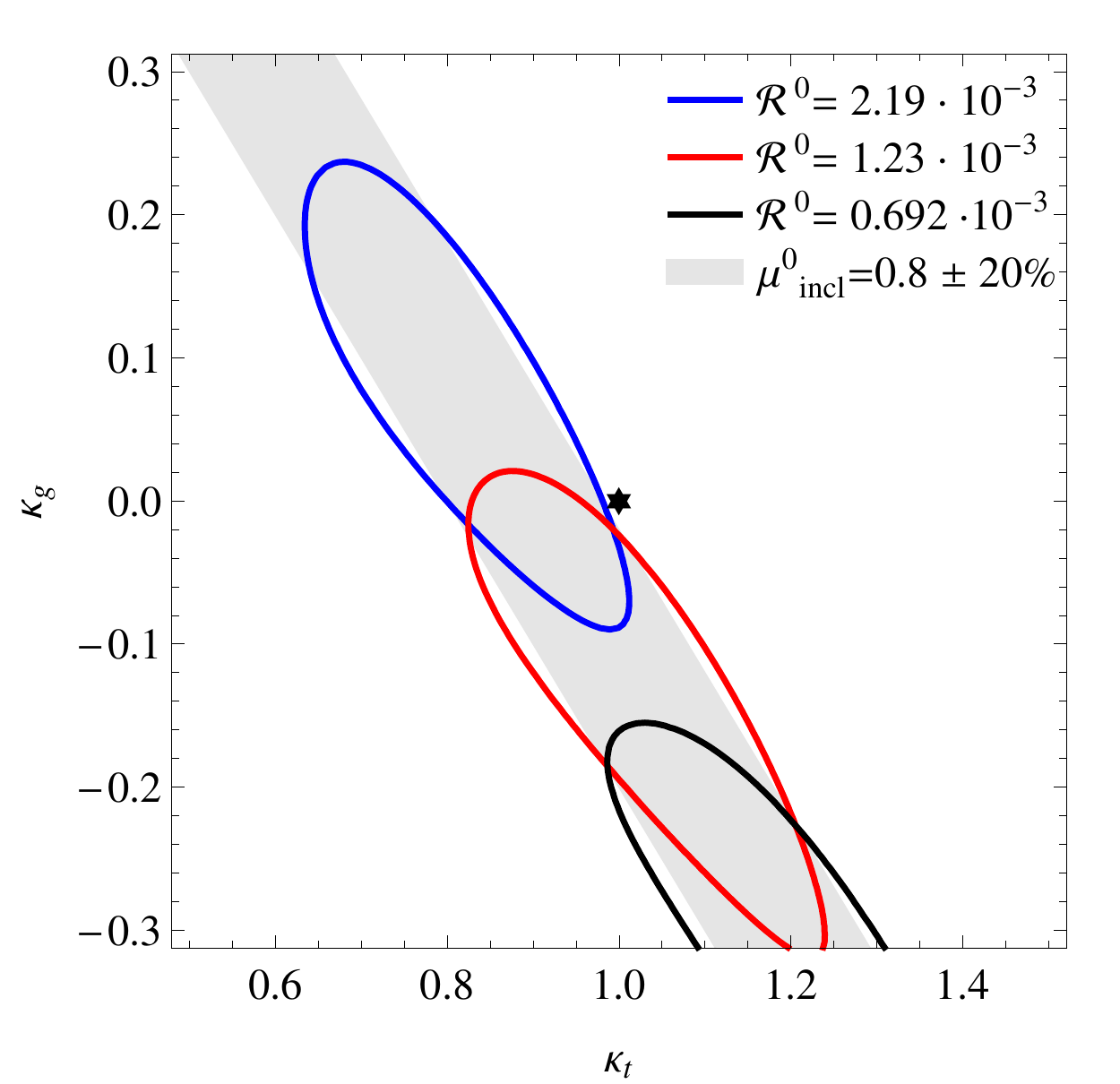}
    \label{fig:mui08}
  } \subfigure[\(\mu^{0}_\tn{incl}=1.0\)]{
    \includegraphics[width=0.45\linewidth]{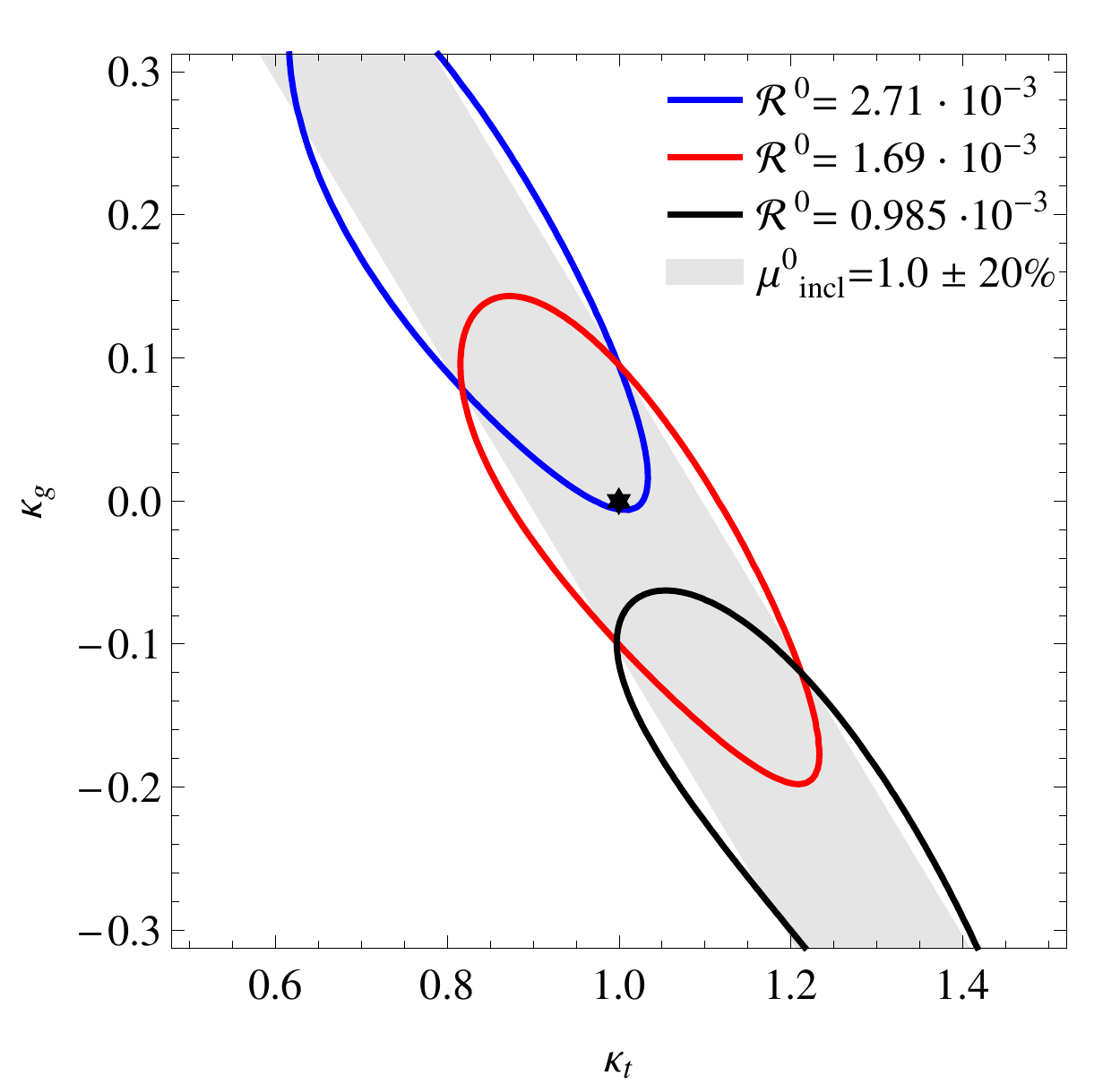}
    \label{fig:mui10}
  } \subfigure[\(\mu^{0}_\tn{incl}=1.2\)]{
    \includegraphics[width=0.45\linewidth]{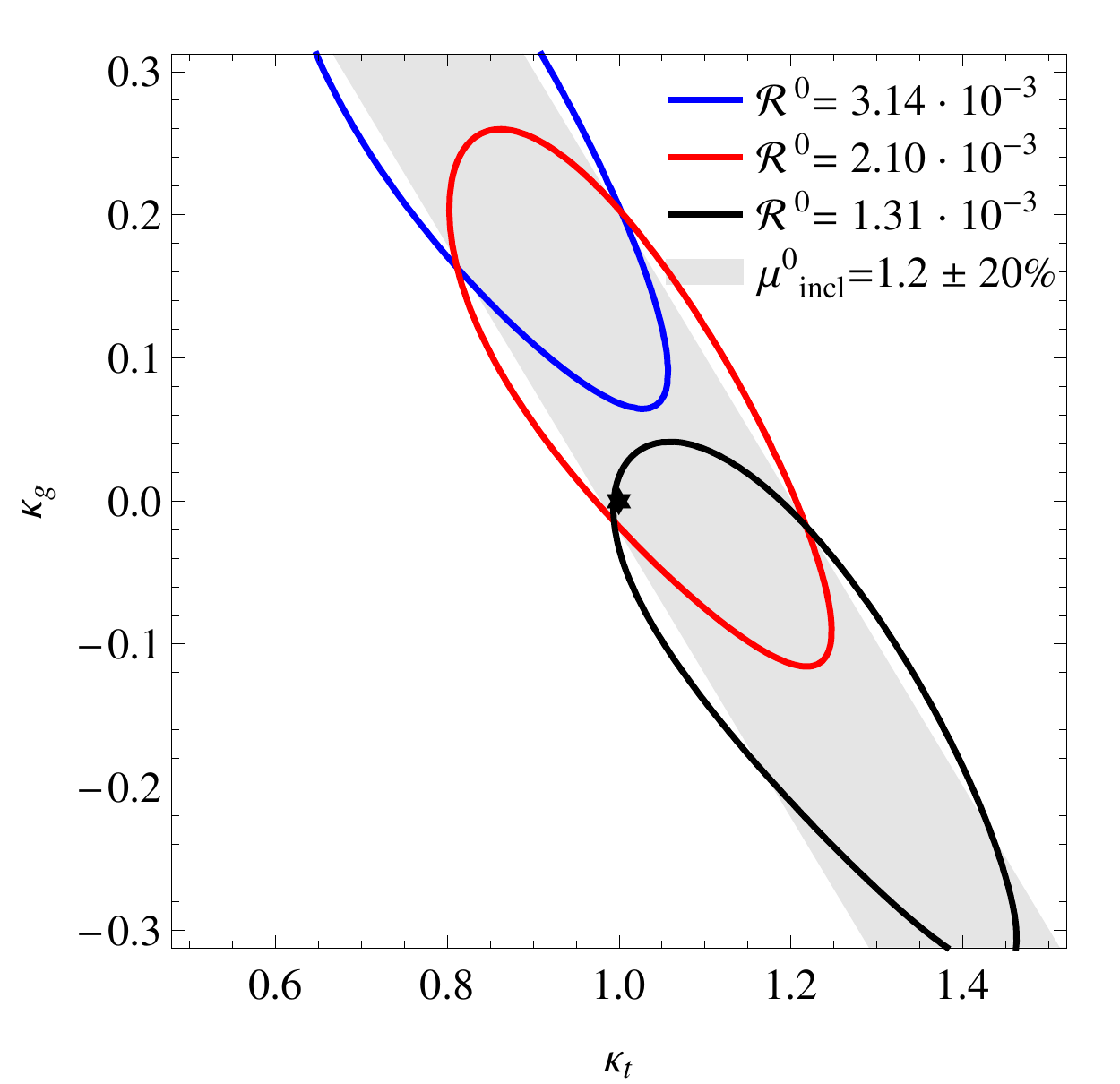}
    \label{fig:mui12}
  } \subfigure[Scale variation]{
    \includegraphics[width=0.45\linewidth]{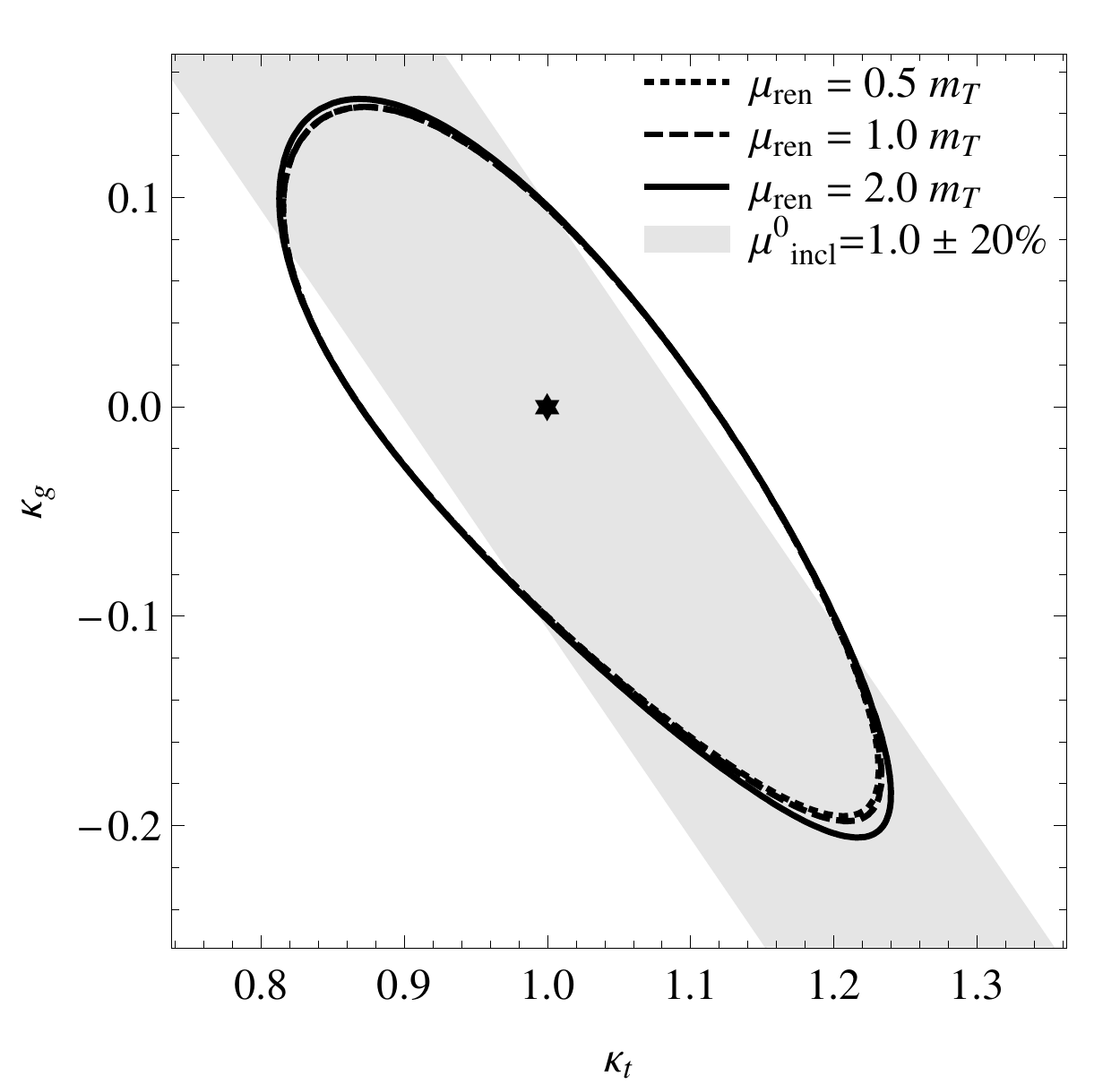}
    \label{fig:scale}
  }
  \caption[]{Figures \subref{fig:mui08}-\subref{fig:mui12} show the $95\%$ CL contours obtained from the $\chi^{2}$ in Eq.~\eqref{eq:1} for different choices of the actual parameters $\kappa_{t}^{0}$ and $\kappa_{g}^{0}$, or equivalently of $\mu_{\tn{incl}}^{0}$ and $\mathcal{R}^{0}$. The colors blue, red and black correspond to $\kappa_t^0=0.8$, $1.0$ and $1.2$, respectively, or equivalently to the indicated values of $\mathcal{R}^{0}=\mathcal{R}(\kappa_{t}^{0},\sqrt{\mu_{\mathrm{incl}}^{0}}-\kappa_{t}^{0})$. The gray band is obtained by considering only the inclusive measurement. The SM point is indicated by the black star. Figure~\subref{fig:scale} shows the variation of the $95\%$ CL contours for different choices of the renormalization and factorization scale $\mu$. For all plots we assumed an integrated luminosity of \(\int\mathcal{L}\,dt =3\,\tn{ab}^{-1}\) and \(\sqrt{s}=14\,\tn{TeV}\).}
  \label{fig:contourplots}
\end{figure}

We assign a $10\%$ systematic uncertainty on each of the cross sections that appear in Eq.~\eqref{eq:4}, and in addition we consider statistical uncertainties on the number of signal events $N_{p_T^{\mathrm{min}}}=\sigma_{p_T^{\mathrm{min}}}(\kappa_t,\kappa_g)\, K_{p_T^{\mathrm{min}}} \,\epsilon_{\mathrm{tot}} \,\int \mathcal{L}\,dt\,$, where $\int \mathcal{L}\,dt = 3\,\mathrm{ab}^{-1}$ is the integrated LHC luminosity. We do not consider any backgrounds in our exploratory study. The two observables $\mu_{\tn{incl}}$ and $\mathcal{R}$ are combined via a simple $\chi^2$,
\begin{equation}
  \label{eq:1}
  \chi^2(\kappa_t,\kappa_g)=\left(\frac{\mu_{\tn{incl}}(\kappa_t, \kappa_g)-\mu_{\tn{incl}}^{0}}{\delta\mu_{\tn{incl}}}\right)^2+\left(\frac{\mathcal{R}(\kappa_t,\kappa_g)-\mathcal{R}^{0}}{\delta\mathcal{R}}\right)^2\,,
\end{equation}
where
\begin{equation}
  \label{eq:11}
\frac{\delta\mu_{\mathrm{incl}}}{\mu_{\mathrm{incl}}^{0}}=\delta_{\mathrm{sys}}\,,\qquad   \frac{\delta\mathcal{R}}{\mathcal{R}^{0}}\,=\,\sqrt{N_{150\,\mathrm{GeV}}^{-1}+N_{650\,\mathrm{GeV}}^{-1}+2 \delta_{\mathrm{sys}}^{\,2}}\,,
\end{equation}
with $\delta_{\mathrm{sys}}=0.1\,$. The potential of the boosted Higgs measurement to break the degeneracy along the $\,\kappa_{t}+\kappa_{g}\,=\,\mathrm{constant}\,$ direction is shown in Figs.~\ref{fig:mui08}-\subref{fig:scale}. We consider three different assumptions on the observed inclusive signal strength, $\mu^{0}_{\mathrm{incl}}=0.8,1,1.2\,$, and on the actual value of the $h\bar{t}t$ coupling, $\kappa_{t}^{0}=0.8,1,1.2\,$. The actual value of $\kappa_{g}$ is then fixed by $\kappa_{g}^{0}=\sqrt{\mu_{\mathrm{incl}}^{0}}-\kappa_{t}^{0}$, and $\mathcal{R}^{0}=\mathcal{R}(\kappa_{t}^{0},\kappa_{g}^{0})$. In Figs.~\ref{fig:mui08}-\subref{fig:mui12} we show the $95\%$ CL contours obtained from the $\chi^{2}$ in Eq.~\eqref{eq:1}, assuming an integrated luminosity of $3\,\mathrm{ab}^{-1}$ at the 14\,TeV LHC. The blue, red and black contours correspond to $\kappa_{t}^{0}=0.8,1,1.2\,$, respectively. The error band obtained considering only inclusive production is shown as well, shaded in gray. It can be clearly seen that including the boosted Higgs measurement allows one to break the degeneracy: for example, assuming a standard inclusive rate (\emph{i.e.} $\mu_{\mathrm{incl}}^{0}=1,$ Fig.~\ref{fig:mui10}) but a $h\bar{t}t$ coupling deviating by $\pm 20\%$ from the standard value (black and blue curves, respectively), the SM point can be excluded at approximately $95\%$ CL. Figure~\ref{fig:scale} shows instead the sensitivity of our results to a variation of the factorization and renormalization scales. The SM is assumed as input ($\kappa_t^0=1$ and $\kappa_g^0=0$) and the scale $\mu = \mu_{\tn{ren}}=\mu_{\mathrm{fact}}$ is set to \(m_T/2\), \(m_T\) and \(2 m_T\). Because we employ the ratio of cross sections $\mathcal{R}$, the preferred region in the plane of couplings depends only mildly on the scale choice. 
\clearpage
\subsection{Limitations and possible improvements}
Firstly, we should make it clear that the above is only a rough first estimate of the resolving power of the boosted Higgs regime, based purely on signal rates and without considering any backgrounds. Furthermore, the assumption of a $10\%$ systematic uncertainty on the boosted Higgs cross sections is likely to be optimistic. Secondly, while in this paper we focused on $h\to \tau\tau\,$, we wish to mention briefly other \emph{a priori} interesting choices for the Higgs decay mode. The dominant channel $h\to b\bar{b}$ seems challenging due to the overwhelming QCD backgrounds. At very large $p_{T}$ the $b$-quarks are collimated, leading in practice to a dijet-like topology; this could provide an interesting application for jet substructure techniques. On the other hand, the decay $h\to WW$ should provide an attractive alternative to $h\to \tau\tau\,$. These issues certainly warrant a more detailed study. 
%
%
\section{Resolving the top partner spectrum in composite Higgs models}
\label{sec:MCHM}
%
The degeneracy between the couplings $\kappa_{t}$ and $\kappa_{g}$ assumes a peculiar form in composite Higgs models. It turns out that in the most popular realizations of the Higgs as a composite pseudo-Nambu--Goldstone (pNGB) boson, the combination $\kappa_{t}+\kappa_{g}$ is insensitive to the `top partners' (the fermionic resonances that are expected to accompany the top)~\cite{Falkowski:2007hz,Low:2010mr,Azatov:2011qy, Delaunay:2013iia, Montull:2013mla}, as a consequence of an exact cancellation among their contribution to $\kappa_{t}$ and to $\kappa_{g}$. It follows that no indirect information on the resonances can be extracted from the inclusive Higgs rates. To review how this result arises in the well-known `Minimal Composite Higgs Model (MCHM)', based on the coset $SO(5)/SO(4)$, we first note that\footnote{This formula holds in the natural operator basis for a non-linear $\sigma$-model, see for example Ref.~\cite{Low:2010mr} for a detailed discussion. The cancellation of the effects of top partners is physical and therefore independent of the basis.}
\begin{equation} \label{hgg ch}
\kappa_{t}+\kappa_{g}=v\left(\frac{\partial}{\partial h}\log \det\mathcal{M}_{t}(h)\right)_{\left\langle h\right\rangle}\,,
\end{equation}    
where $\mathcal{M}_{t}(h)$ is the mass matrix in the top sector. The partial compositeness structure and the pNGB nature of the Higgs imply~\cite{Montull:2013mla} that the determinant factorizes as $\det \mathcal{M}_{t}(h)=m_{t}^{0}(h)\times \det M_{c}\,$, where $m_{t}^{0}$ denotes the top mass,\footnote{More precisely, $m_{t}^{0}$ is the expression of the top mass when corrections to the wavefunctions of $t_{L,R}$ due to the mixing with top partners are ignored~\cite{Montull:2013mla}.} whereas $M_{c}$ is the mass matrix of top partners. We are assuming to be in the field basis where non-derivative interactions of the Goldstone bosons appear only in the linear mixing terms, see Eq.~\eqref{eq:mchm5} later for an example. Therefore the matrix $M_{c}$ does not depend on $h\,$, and its determinant drops out of Eq.~\eqref{hgg ch}. In addition, by means of a spurion argument~\cite{Azatov:2011qy} it can be shown that in models where only one $SO(4)$ invariant $I_{LR}(h/f)$ can be built out of the embeddings of $t_{L}$ and $t_{R}$, one has $m_{t}^{0}(h)\propto I_{LR}(h/f)$, with $f$ denoting the decay constant of the non-linear sigma model. Thus Eq.~\eqref{hgg ch} yields simply $\kappa_{t}+\kappa_{g} = f_{g}(\xi\equiv v^{2}/f^{2})$, where $f_{g}$ is some function satisfying $f_{g}(\xi\to 0) = 1\,$. Interestingly, the most viable (and popular) realizations of the MCHM feature only one left-right invariant: for example, the choices $\mathbf{5}_{L}+\mathbf{5}_{R}\,$, $\mathbf{10}_{L}+\mathbf{10}_{R}$ and $\mathbf{14}_{L}+\mathbf{1}_{R}\,$ all lead to $I_{LR}=\sin(2h/f)$ and therefore to $\kappa_{t}+\kappa_{g}=(1-2\xi)/\sqrt{1-\xi}\,$. Furthermore, the insensitivity of the low-energy $hgg$ coupling to the resonances holds also in several Little Higgs models~\cite{Low:2010mr}.\footnote{Exceptions to this result exist. In particular, the insensitivity to the resonance masses does not hold in versions of the MCHM where multiple left-right invariants appear, such as MCHM$_{14}$, where $q_{L}$ and $t_{R}$ are embedded into the $\mathbf{14}$-dimensional representation of $SO(5)$.}\\

\subsection{A simple explicit model}

\begin{figure}[tb]
  \centering
  \includegraphics[width=0.6\linewidth]{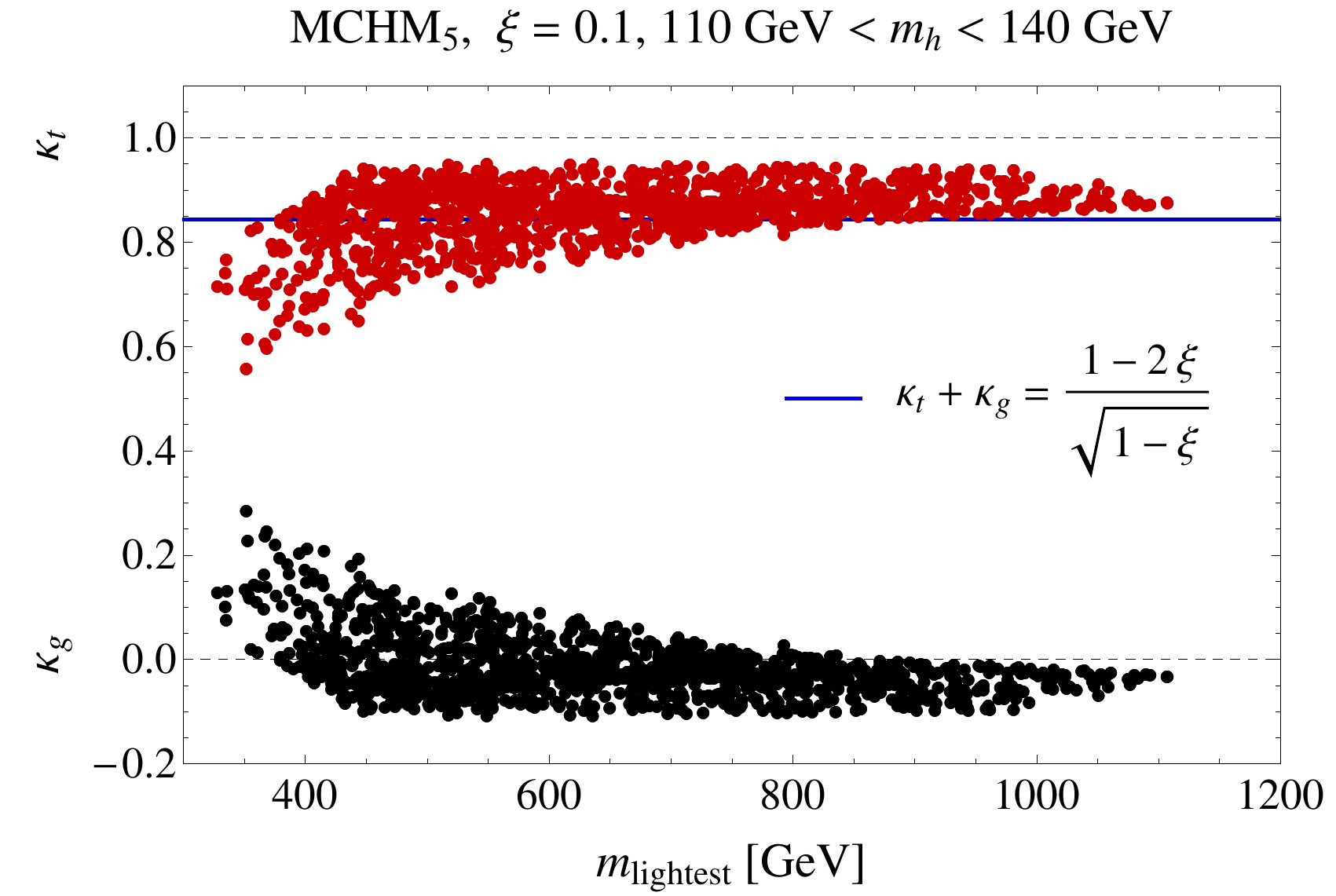}
  \caption{Distribution of the couplings $\kappa_{t}$ and $\kappa_{g}$ versus the mass of the lightest top partner, as obtained from the numerical scan in the two-site MCHM$_{5}$.}
  \label{fig:comp}
\end{figure}

With the twofold goal of making a simple, concrete example of the above argument and of quantifying the expected size of $\kappa_{g}$ in realistic composite Higgs models, we consider a `two-site' version of the MCHM$_{5}$~\cite{Panico:2011pw}, where one complete $SO(5)$ multiplet of composite fermions is introduced: the fermion Lagrangian reads
\begin{equation} \label{eq:mchm5}
\mathcal{L}_{f}= i\overline{q}_{L}\slashed{D}q_{L}+i\overline{t}_{R}\slashed{D}t_{R} +i \overline{\psi}\slashed{D}\psi -m_{4}\overline{\psi}_{4}\psi_{4}-m_{1}\overline{\psi}_{1}\psi_{1}-(\lambda_{q}\overline{\mathcal{Q}}_{L}U^{T}\psi_{R}+\lambda_{u}\overline{\psi}_{L}U\mathcal{Q}_{R} + \mathrm{h.c.})\,,
\end{equation}
where $\psi=(\psi_{4},\,\psi_{1})^{T}$ is a complete $\mathbf{5}$ of composite fermions, which decomposes as $\mathbf{5}\sim \mathbf{4}\oplus \mathbf{1}$ under $SO(4)$, and $\mathcal{Q}_{L,R}$ are the embeddings into incomplete $\mathbf{5}\,$s of $q_{L}$ and $t_{R}$, respectively. The matrix $U$ contains the Goldstone bosons, and the associated decay constant is labeled by $f$\,. In this setup, the Higgs potential arising from top loops is partially calculable, having the form $V\simeq \alpha \sin^{2}(h/f) + \beta \sin^{4}(h/f)$, where $\alpha$ is logarithmically divergent whereas $\beta$ is finite. It follows that the divergence in $\alpha$ can be absorbed by adding a suitable counterterm and fixing the value of the Higgs VEV $v$ as renormalization condition, while the Higgs mass can still be predicted~\cite{Matsedonskyi:2012ym}. In order to have a qualitative picture, we set $\xi = 0.1$ and scan the parameter space of the model requiring $m_{t} = 150$\,GeV (corresponding roughly to the running top mass at scale $\sim \mathrm{TeV}$) and $110\,\mathrm{GeV} < m_{h} < 140\,\mathrm{GeV}$. Figure~\ref{fig:comp} shows the distribution of the couplings $\kappa_{t}$ and $\kappa_{g}$ versus the mass of the lightest top partner, as obtained from the numerical scan. It is immediate to see `by eye' that, as predicted by the argument summarized above, the sum $\kappa_{t}+\kappa_{g}$ is the same for all points, being given by $(1-2\xi)/\sqrt{1-\xi}\,$.     
Assuming for example a large $t_{R}$ compositeness, $\kappa_{g}$ has the expression (see Ref.~\cite{Delaunay:2013iia} for the complete formula)
\begin{equation} \label{eq:kg comp}
\kappa_{g}= \xi\,\sin^{2}\theta_{R}\left(\frac{m_{1}^{2}-m_{4}^{2}}{m_{4}^{2}}\right)+O(\sin^{2}\theta_{L})\,,
\end{equation}
where $\sin\theta_{L,R}=\lambda_{q,u}/\sqrt{m_{4,1}^{2}+\lambda_{q,u}^{2}}$ measures the degree of compositeness of $t_{L,R}\,$. Combining the full analytical expression of $\kappa_{g}$~\cite{Delaunay:2013iia} with the approximate formula for the Higgs mass~\cite{Matsedonskyi:2012ym} 
\begin{equation}
m_{h}^{2}\simeq \frac{N_{c}}{\pi^{2}}\frac{m_{t}^{2}}{f^{2}}\frac{m_{T}^{2}m_{\tilde{T}}^{2}}{m_{T}^{2}-m_{\tilde{T}}^{2}}\log\frac{m_{T}^{2}}{m_{\tilde{T}}^{2}}\,\qquad \left(m_{T,\tilde{T}}=\sqrt{m_{4,1}^{2}+\lambda_{q,u}^{2}}\right)\,,
\end{equation}
we find that `large' values of $\kappa_{g}$, $\left|\kappa_{g}\right| > \xi\,$, are possible only in the presence of a very light resonance, $m_{\mathrm{lightest}}\ll 1\,\mathrm{TeV}$. This is confirmed by the numerical scan: the points yielding $\left|\kappa_{g}\right| > \xi\,$ have a resonance lighter than $500\,\mathrm{GeV}$.

Combining the above discussion with the results of Section~\ref{sec:analysis}, and in particular with Fig.~\ref{fig:contourplots}, it is clear that boosted Higgs production can reveal the presence of light top partners, which would remain otherwise `hidden' in the measurement of the inclusive rate.\footnote{See Ref.~\cite{Banfi:2013yoa} for a recent study of $pp\to h+\tn{jet}$ within the context of composite Higgs models.} Thus the large-$p_{T}$ Higgs production can usefully complement the information coming from direct searches for the fermionic resonances.

\subsection{Validity of the effective theory approach}
\begin{figure}[tb]
  \begin{center}
  \hspace{-2mm}\includegraphics[width=0.49\linewidth]{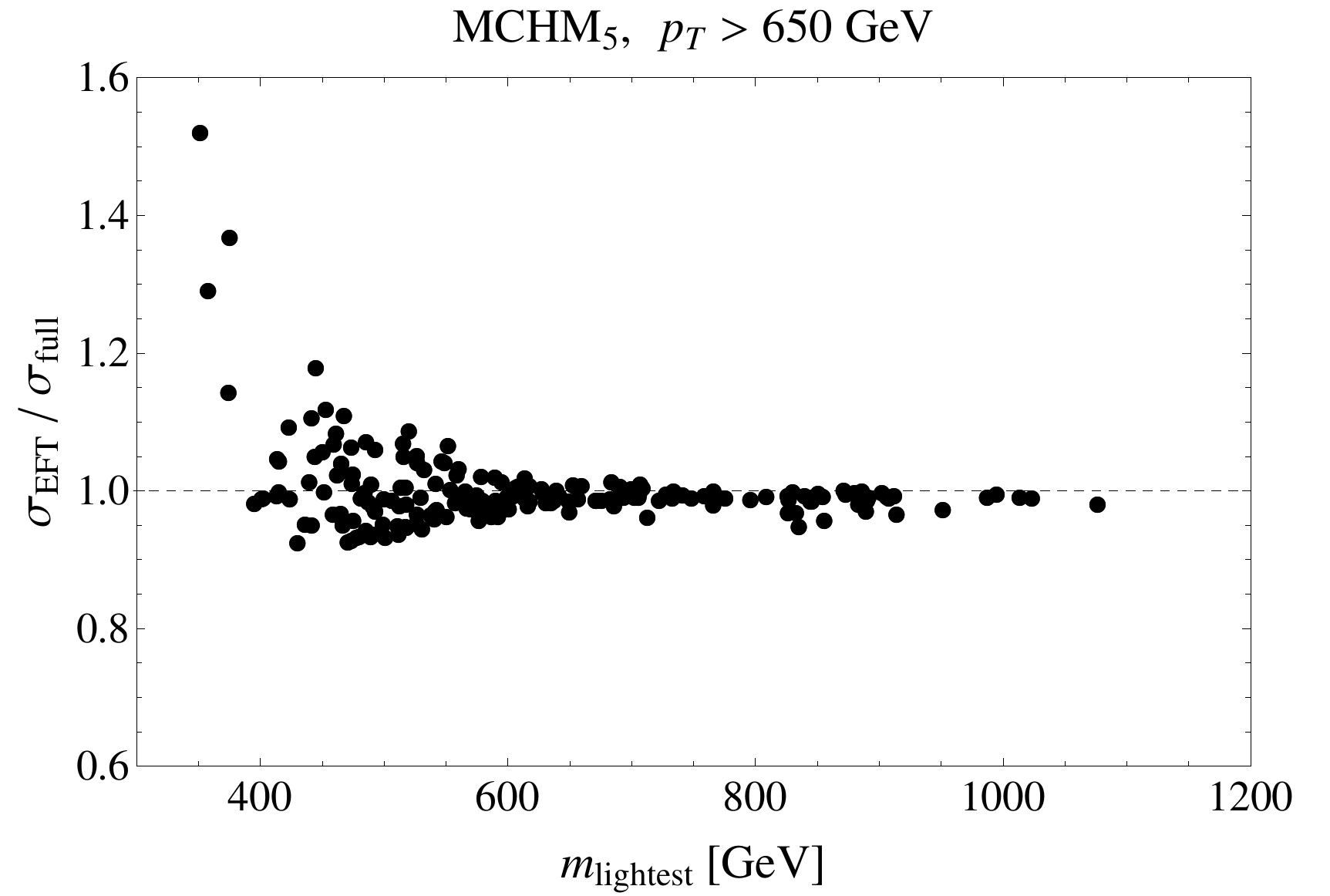}\hspace{2mm}
\centering
  \includegraphics[width=0.48\linewidth]{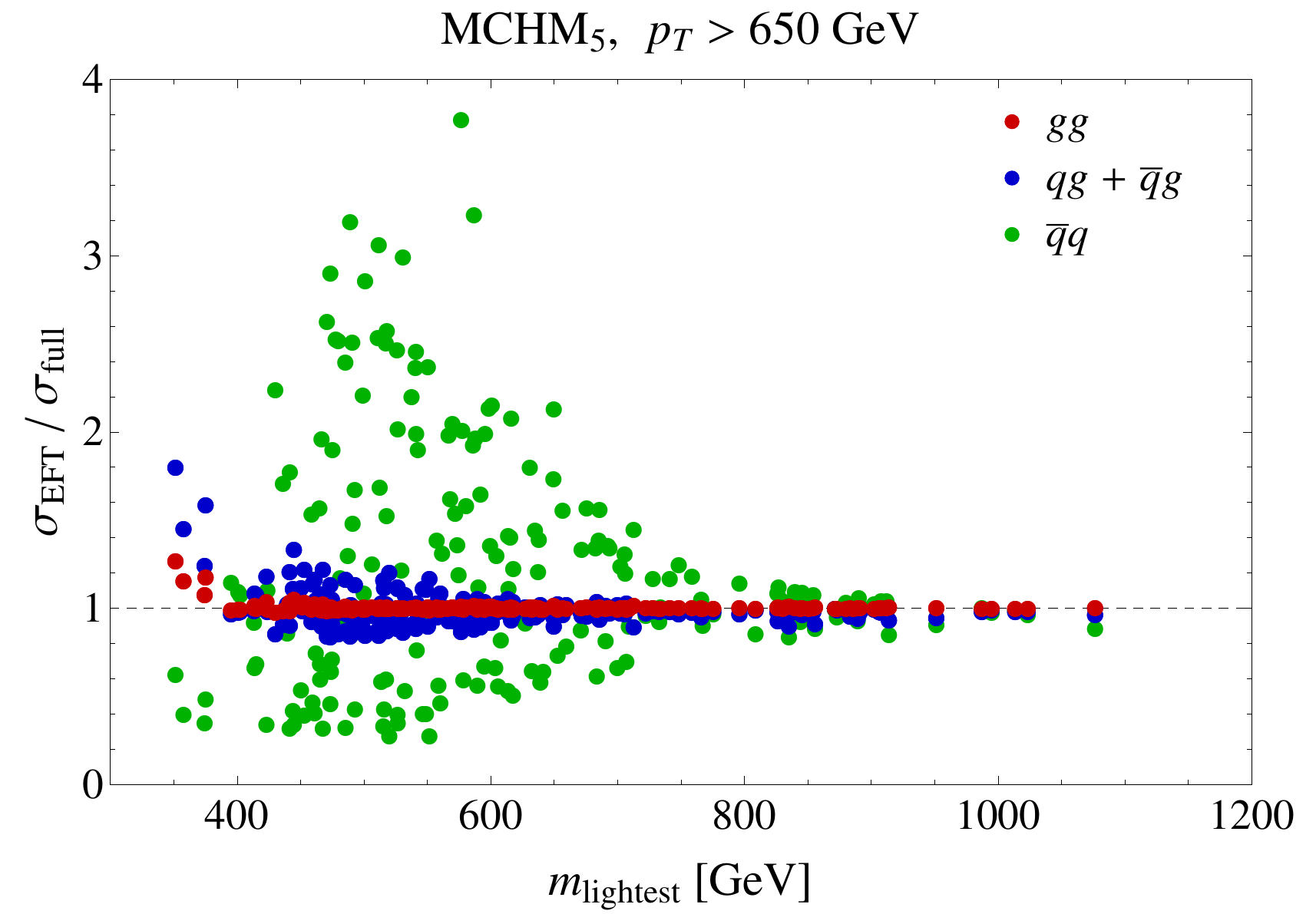}
  \caption{Ratio of the boosted Higgs cross section computed within the effective theory to the exact cross section computed retaining the complete form factors, versus the mass of the lightest top partner, for a sample set of points in the parameter space of MCHM$_{5}$. A transverse momentum cut $p_{T}>650\,\mathrm{GeV}$ is applied. The left panel shows the total cross section $pp\to h+\tn{jet}\,$, whereas the right panel shows the three partonic channels $gg, qg, q\bar{q} \to h+\tn{jet}$ individually.}
  \label{fig:EFT}
  \end{center}
\end{figure}

In our analysis of the MCHM$_{5}$ we have assumed the validity of the Effective Field Theory (EFT) description of new physics states, and computed the coefficient $\kappa_{g}$ by integrating out the top partners. However, given a transverse momentum cut $p_{T}>p_{T}^{\mathrm{min}}$, we naively expect the EFT to break down if there is at least one resonance with mass $M<p_{T}^{\mathrm{min}}\,$. Since we defined the very boosted region by $p_{T}^{\mathrm{min}}=650\,\mathrm{GeV}$, we expect the EFT to be inaccurate for the spectra which feature a top partner with $M\lesssim 650\,\mathrm{GeV}$. This is quantified in Fig.~\ref{fig:EFT}, where we compare the boosted cross section computed in the EFT approximation to the full result obtained by keeping the complete loop form factors. We find that the EFT is accurate within $10\%$ for top partners as light as $500\,\mathrm{GeV}$. We also show a comparison of the EFT and exact cross sections for each partonic initial state, namely for $gg, qg$ and $q\bar{q}\,$. For the $gg$ channel the EFT is accurate within a few percent, whereas for $qg$ the approximation works within $20\%$ for resonances above $500\,\mathrm{GeV}$. Recalling from Table~\ref{tab:coeff} that for $p_{T}>650\,\mathrm{GeV}$ the $gg$ and $qg$ channels each make up about $50\%$ of the total cross section, this yields the already mentioned $10\%$ overall accuracy. By contrast, the breakdown of the EFT in the $q\bar{q}$ channel is striking, as already noticed in Ref.~\cite{Harlander:2012hf}. This effect is however numerically small, since the $q\bar{q}$ channel only contributes about $1\div 2\%$ of the total cross section.  

%

\section{Resolving the stop spectrum/mixing in supersymmetric models}
\label{sec:SUSY}

\begin{figure}[]
  \centering
  \begin{tikzpicture}[baseline=(current bounding box.center), line width=1 pt, scale=1.7]
    \begin{scope}[shift={(-2.5,0)}]%
      \draw[scalarnoarrow] (0:0.5)--(72:0.5);%
      \draw[scalarnoarrow] (72:0.5)--(144:0.5);%
      \draw[scalarnoarrow] (144:0.5)--(216:0.5);%
      \draw[scalarnoarrow] (216:0.5)--(288:0.5);%
      \draw[scalarnoarrow] (288:0.5)--(360:0.5);%
      
      \draw[gluon] (-1.5,0.5)--(144:0.5); %
      \draw[gluon] (-1.5,-0.5)--(216:0.5); %
      \draw[gluon] (72:0.5)--(1.5,0.5); %
      \draw[scalar] (288:0.5)--(1.5,-0.5); %
      \draw[scalarnoarrow] (0:0.5)--(0:1.0); %
      
      \begin{scope}[shift={(0:1.0)}]
        \draw (45:.1) -- (-135:.1);%
        \draw (-45:.1) -- (135:.1);%
      \end{scope}
      \begin{scope}[shift={(0.5,0)}]
        \draw[color=black, fill=white] (0,0) circle (0.1);%
        \draw (45:.1) -- (-135:.1);%
        \draw (-45:.1) -- (135:.1);%
      \end{scope}
      \begin{scope}[shift={(288:0.5)}]
        \draw[color=black, fill=white] (0,0) circle (0.1);%
        \draw (45:.1) -- (-135:.1);%
        \draw (-45:.1) -- (135:.1);%
      \end{scope}

      \node at (0.0,0.0) {\(\tilde{t},\tilde{b}\)}; %
      \node at (26:0.6) {\(L\)};%
      \node at (108:0.6) {\(L\)};%
      \node at (180:0.6) {\(L\)};%
      \node at (252:0.6) {\(L\)};%
      \node at (325:0.55) {\(R\)};%

      \node at (-1.7,0.5) {\(g\)}; %
      \node at (1.7,0.5) {\(g\)}; %
      \node at (-1.7,-0.5) {\(g\)}; %
      \node at (1.7,-0.5) {\(h\)}; %

    \end{scope}
%
    \begin{scope}[shift={(2.5,0)}]
      \draw[scalarnoarrow] (0:0.5)--(90:0.5);%
      \draw[scalarnoarrow] (90:0.5)--(180:0.5);%
      \draw[scalarnoarrow] (180:0.5)--(270:0.5);%
      \draw[scalarnoarrow] (270:0.5)--(360:0.5);%
      \draw[scalarnoarrow] (0:0.5)--(0:1.0);%
      
      \draw[gluon] (-1.5,0.5)--(90:0.5); %
      \draw[gluon] (-1.5,-0.5)--(180:0.5); %
      \draw[gluon] (90:0.5)--(1.5,0.5); %
      \draw[scalar] (270:0.5)--(1.5,-0.5); %

      \begin{scope}[shift={(0:1.0)}]%
        \draw (45:.1) -- (-135:.1);%
        \draw (-45:.1) -- (135:.1);%
      \end{scope}
      \begin{scope}[shift={(270:0.5)}]%
        \draw[color=black, fill=white] (0,0) circle (0.1);%
        \draw (45:.1) -- (-135:.1);%
        \draw (-45:.1) -- (135:.1);%
      \end{scope}
      \begin{scope}[shift={(0.5,0)}]%
        \draw[color=black, fill=white] (0,0) circle (0.1);%
        \draw (45:.1) -- (-135:.1);%
        \draw (-45:.1) -- (135:.1);%
      \end{scope}
      
      \node at (0.0,0.0) {\(\tilde{t},\tilde{b}\)}; %
      \node at (30:0.5) {\(L\)};%
      \node at (148:0.5) {\(L\)};%
      \node at (-148:0.5) {\(L\)};%
      \node at (320:0.5) {\(R\)};%

      \node at (-1.7,0.5) {\(g\)}; %
      \node at (1.7,0.5) {\(g\)}; %
      \node at (-1.7,-0.5) {\(g\)}; %
      \node at (1.7,-0.5) {\(h\)}; %
    \end{scope}

    \begin{scope}[shift={(-2.50,-2)}]%
      \node at (-1.7,0.5) {\(g\)}; %
      \node at (-1.7,-0.5) {\(g\)}; %
      \node at (1.7,0.5) {\(g\)}; %
      \node at (1.7,-0.5) {\(h\)}; %

      \draw[gluon] (-1.5,0.5)--(-.3,0.3);%
      \draw[gluon] (-1.5,-0.5)--(-.3,-0.3);%
      \draw[gluon] (-.3,0.3)--(1.5,0.5);%
      \draw[gluon] (-.3,0.3)--(-.3,-0.3);%
      
      \draw[scalarnoarrow] (-0.3,-.3) arc (133:30:0.85);%
      \draw[scalarnoarrow] (-0.3,-.3) arc (-150:-47:.85);%
      \draw[scalar] (1.0,-0.5)--(1.5,-0.5);%

      \node at (0.3,-0.4) {\(\tilde{t},\tilde{b}\)};%
    \end{scope}%

    \begin{scope}[shift={(2.50,-2)}]%
      \node at (-1.7,0.5) {\(q\)}; %
      \node at (-1.7,-0.5) {\(g\)}; %
      \node at (1.7,0.5) {\(q\)}; %
      \node at (1.7,-0.5) {\(h\)}; %

      \draw[fermion] (-1.5,0.5)--(-.3,0.3);%
      \draw[gluon] (-1.5,-0.5)--(-.3,-0.3);%
      \draw[fermion] (-.3,0.3)--(1.5,0.5);%
      \draw[gluon] (-.3,0.3)--(-.3,-0.3);%
      
      \draw[scalarnoarrow] (-0.3,-.3) arc (133:30:0.85);%
      \draw[scalarnoarrow] (-0.3,-.3) arc (-150:-47:.85);%
      \draw[scalar] (1.0,-0.5)--(1.5,-0.5);%

      \node at (0.3,-0.4) {\(\tilde{t},\tilde{b}\)};%
    \end{scope}%
  \end{tikzpicture}
  \caption{Example Feynman diagrams for \(pp\to h+\tn{jet}\) involving supersymmetric particles. In addition, there are diagrams like those in Fig.~\ref{fig:SMdiag}, but with the quarks in the loops replaced by squarks.}
  \label{fig:SUSYdiag}
\end{figure}
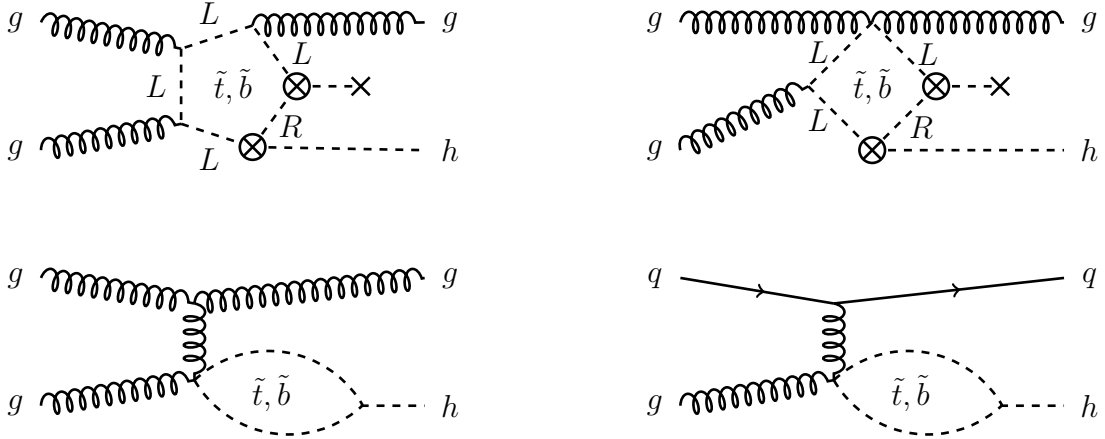

\begin{figure}[tb]
  \centering
  \includegraphics[width=1\linewidth]{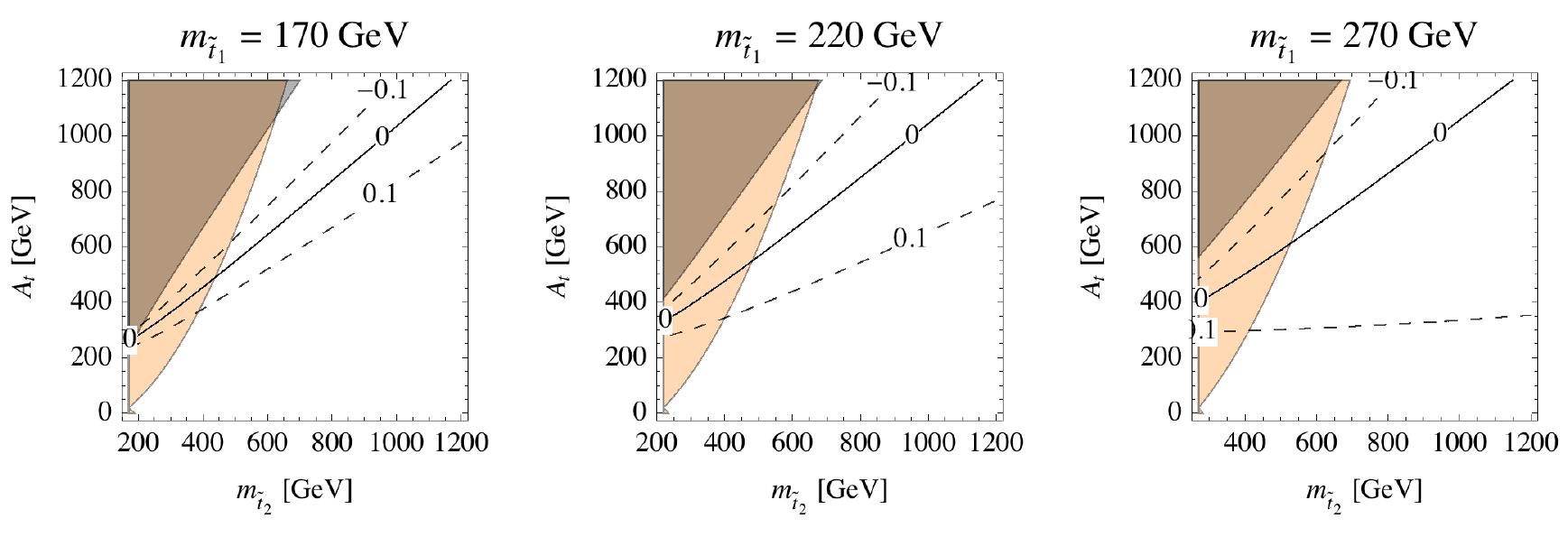}
  \caption{We show the relative deviation of the $gg\to h$ amplitude from the SM value due to the
    stop contribution, $\Delta_t =0, \pm 0.1 $ (solid and dashed lines, respectively). In addition
    we plot the parameter space excluded because the soft masses are not real (orange) and an estimate of the vacuum stability constraint (grey).}
  \label{fig:stopvsat}
\end{figure}

In supersymmetric models the $gg\to h$ rate is sensitive to loop contributions of stops and sbottoms, affecting both the
inclusive~\cite{Kunszt:1991qe} and the boosted cross section~\cite{Djouadi:1991tka}. In the following we focus on the Minimal Supersymmetric SM (MSSM) and assume that additional $D$- or $F$-term contributions help lift the Higgs mass into the phenomenologically allowed range. We further assume that the associated beyond-the-MSSM degrees of freedom do not significantly change the properties of the lightest $CP$-even Higgs. Besides the supersymmetric partners of the SM diagrams of Fig.~\ref{fig:SMdiag}, new topologies sensitive to the stop trilinear term $A_t$ can contribute to the cross section, see Fig.~\ref{fig:SUSYdiag}. 
The inclusive signal strength is given by~\cite{Haber:1984zu}
\begin{equation}
  \label{eq:5}
  \frac{\Gamma(g g \to h)}{\Gamma(g g \to h)_{\mathrm{SM}}}=(1+\Delta_t)^2,
\end{equation}
with
\begin{equation}
  \label{eq:6}
  \Delta_t\approx \frac{m_t^2}{4}\left(\frac{1}{m_{\tilde{t}_1}^2}+\frac{1}{m_{\tilde{t}_2}^2}-\frac{(A_t-\mu/\tan\beta)^{2}}{m_{\tilde{t}_1}^2m_{\tilde{t}_2}^2}\right),
\end{equation}
We work in the limit where the pseudoscalar Higgs decouples, see \emph{e.g.} Refs.~\cite{Espinosa:2012in,Delgado:2012eu}, and we have ignored small $D$-term contributions. As in composite Higgs models, we encounter a flat direction: for large enough $A_t$, the stop contribution to the inclusive rate can be made to vanish. In Fig.~\ref{fig:stopvsat} we show this cancellation as a function of the stop masses and of $A_t$. This condition requires large $A_t$ and one might worry about vacuum stability, see for instance Ref.~\cite{Reece:2012gi}. A large $A_t$ leads to a large trilinear scalar coupling $\propto h A_t \tilde t_L \tilde t_R^*$. If all three fields aquire vacuum expectation values, the potential can have a deep charge- and color-breaking minimum, separated only by a relatively low potential barrier from the usual electroweak vacuum. A rough but conservative estimate of the vacuum stability condition is given by~\cite{Casas:1995pd,Kusenko:1996jn}
\begin{equation}
  \label{eq:vacuum_stability}
  A_t^2 + 3 \mu^2 < a\cdot \left(m_{\tilde t_1}^2 + m_{\tilde t_2}^2\right),
\end{equation}
with $a \approx 3$. This vacuum stability
condition is shown in Fig.~\ref{fig:stopvsat}, colored in grey. We further identify the regions of
parameter space which are excluded because the soft masses $M_{Q_3}, M_{U_3}$ are not real (orange). 

Direct limits from ATLAS and CMS significantly constrain the allowed parameter space. An exhaustive
re-analysis of the spectra and decays of all possible light and mixed stops is, however, beyond the
scope of our paper. While current experimental searches exclude a significant part of the stop
parameter space, these limits soften considerably for larger LSP masses, close to kinematic
degeneracies and in the presence of more complicated decay chains, or in the absence of the
traditional missing $E_T$ signatures (see \emph{e.g.}
Refs.~\cite{Kats:2011it,Fan:2012jf,Fan:2011yu,Bai:2013xla}). In particular, light stops with
$m_{\tilde t_1} - m_{\tilde\chi_0} \approx m_t$ are still compatible with data~\cite{ATLAS:stop_searches,CMS:stop_searches}. It is therefore interesting to ask whether we can be sensitive to light and mixed stops independently of the assumptions on their decays and even if their contribution cancels in the inclusive rate. 

\begin{figure}[h!]
  \centering
  \includegraphics[width=0.49\linewidth]{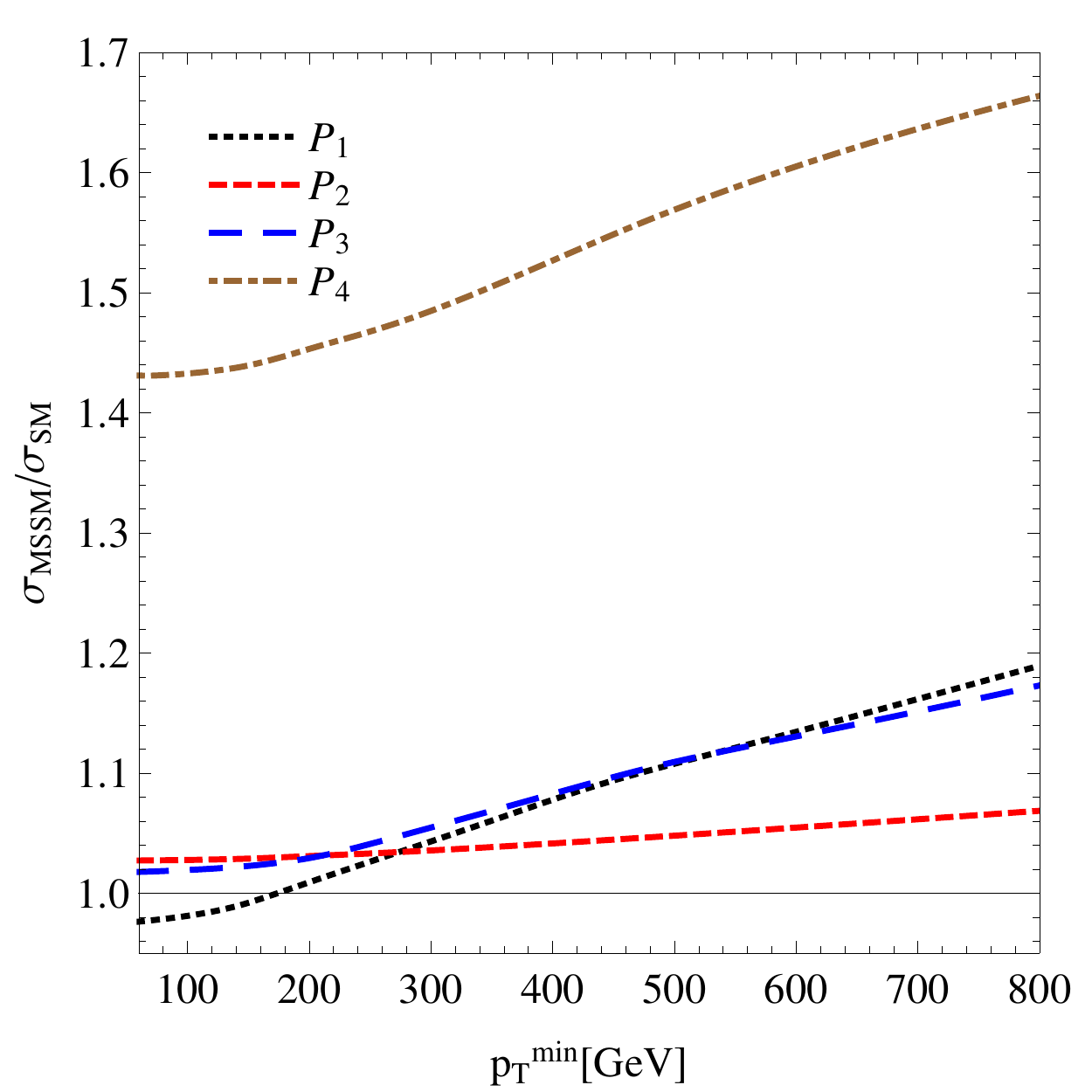}
  \caption{Cross section for boosted Higgs in the MSSM, normalized to the SM value, as a function of the transverse momentum cut $p_{T}^{\mathrm{min}}$. The different lines correspond to the stop masses and values of \(A_t\) given in Table~\ref{tab:susy-points}.}
  \label{fig:susy}
\end{figure}

We calculated the relevant Feynman diagrams involving the stops using FeynArts-3.7~\cite{Hahn:2000kx} and FormCalc-8.0~\cite{Hahn:1998yk}. We made use of the MSTW 2008 LO PDFs \cite{Martin:2009iq} and of LHAPDF-5.8.9~\cite{Whalley:2005nh}. The factorization and renormalization scales were set to the minimum transverse mass determined by the cut on \(p_T\), $\mu=\sqrt{m_{h}^{2}+p_{T}^{\mathrm{min}\,2}}$. In Fig.~\ref{fig:susy} we show the dependence of the boosted cross-section on the minimum $p_T$ of the Higgs. We plot three benchmark points which would lead to a vanishing contribution to the inclusive cross-section ($P_1,P_2,P_3$) and one point ($P_4$) which shares the same stop masses with ($P_3$) but has a vanishing $A_t$. The latter is included to illustrate that the $A_t$-independent terms are mostly responsible for the departure from the point-like approximation. We find that the $p_T$ dependence resolves the cancellation of the supersymmetric contribution to the inclusive rate. Naturally, since we are showing the case where the leading effect is completely canceled, the overall size of the contribution will be small. In the less extreme case of a partial cancellation the high-$p_T$ measurement could be important to retain sensitivity to the stop contribution. 

In conclusion, we find that a precision measurement of the high-$p_T$ Higgs cross-section can break the degeneracy present in the inclusive rate and render a `stealth' stop contribution visible. 

\begin{table}[htp]
\begin{center}
  \begin{tabular}{|r|r|r|r|r|}
    \hline
    Point & \(m_{\tilde{t}_1}\) [GeV] & \(m_{\tilde{t}_2}\) [GeV] & \(A_t\) [GeV] &
    \(\Delta_t\)\\\hline
    \(P_1\) & 171 & 440 & 490 & 0.0026\\
    \(P_2\) & 192 & 1224 & 1220 & 0.013\\
    \(P_3\) & 226 & 484 & 532 & 0.015\\
    \(P_4\) & 226 & 484 & 0 & 0.18 \\\hline
  \end{tabular}
\end{center}
 \caption{The benchmark points shown in Fig.~\ref{fig:susy}. We set \(\tan \beta=10\), \(M_{A^0}=500\,\tn{GeV}\), \(M_2=1000\,\tn{GeV}\), \(\mu=200\,\tn{GeV}\) and all trilinear couplings to a common value \(A_t\). The remaining sfermion masses were set to
\(1 \,\tn{TeV}\) and the mass of the lightest $CP$-even Higgs was set to \(125\,\tn{GeV}\). }
  \label{tab:susy-points}
\end{table}

\clearpage
\section{Conclusions and outlook}
Gluon fusion is the dominant Higgs production mechanism at the LHC. The current measurements cannot resolve the infrared contribution to the amplitude generated by the top loop from the ultraviolet contribution due to possible new physics, such as the top partners needed to render the Higgs naturally light. The `traditional' approach to break this degeneracy is provided by the $pp\to t\bar{t}h$ process, where a direct measurement of the $h\bar{t}t$ coupling can be performed. The objective experimental difficulty of this channel, however, makes it worthwhile to investigate alternatives.

In this paper we considered the production of a boosted Higgs in association with a high-$p_{T}$ jet, where the extra radiation allows one to probe the structure of the top loop and thus to disentangle the IR and UV contributions to the amplitude. The potential of the LHC to resolve the degeneracy was estimated, by using an effective approach to parameterize the new physics. We focused on the decay $h\to \tau\tau$ and performed an exploratory analysis based only on signal rates. 

Subsequently, we discussed the application of our results to two main candidates for a solution to the hierarchy problem, namely partial compositeness at the TeV scale and natural SUSY. In the former case, the degeneracy between the IR and UV amplitudes for $gg\to h$ takes the form of a peculiar insensitivity of the inclusive cross section, $\sigma(pp\to h + X)$, to the presence of light fermionic resonances. In the case of supersymmetry, a large $A_{t}$-term can conspire to make the inclusive rate of SM size even in the presence of light stops. We illustrated how, in both scenarios, the boosted Higgs regime can probe the spectrum of top partners (either fermionic or scalar), and thus complement the information coming from direct LHC searches for the resonances.

As a concluding remark, we emphasize that our estimate of the resolving power of the boosted Higgs regime is based only on signal rates, without the inclusion of any backgrounds. As such, it should be regarded as a first, preliminary attempt. More detailed studies are needed in order to reliably compare the reach of the boosted Higgs channel with the expected sensitivity of $pp\to t\bar{t}h$ and thus to determine the ultimate accuracy attainable by the LHC in the measurement of the coupling of the Higgs to the top quark.   
\\
{\bf Note added:} While this work was nearing completion,\footnote{For early reports on our results, see the talks by MS at the Spring Conference of the German Physical Society (DPG), Dresden, March 2013; by CG at EPS 2013, Stockholm, July 2013; and by AW at SUSY 2013, Trieste, August 2013~\cite{ourtalks}.} an independent study appeared~\cite{Azatov:2013xha} which has also estimated the sensitivity on the $h\bar{t}t$ coupling determination using the boosted $pp \to h+{\rm jet}$ channel, and defined a variable similar to our observable $\mathcal{R}$ to measure the boosted signal.   
\label{sec:conclusions}

\section*{Acknowledgments}

We thank R.~Contino, M.~Grazzini, A.~Juste, F.~Maltoni, A.~Pomarol, G.~Salam, P.~Uwer and L.~Vacavant for insightful discussions. We are grateful to M.~Spira for correspondence about HIGLU. This research has been partly supported by the European Commission under the ERC Advanced Grant 226371 \emph{MassTeV} and the contract PITN-GA-2009-237920 \emph{UNILHC}. C.G.~is  supported by the Spanish Ministry MICINN under contract
FPA2010-17747. E.S.~was supported in part by the US Department of Energy under grant
DE-FG02-91ER40674 and by the European Commission under the ERC Advanced Grant 267985
\emph{DaMeSyFla}. M.S.~is supported by the Joachim Herz Stiftung. The work of A.W.~was supported in
part by the German Science Foundation (DFG) under the Collaborative Research Center (SFB) 676.

\section*{Appendix: amplitudes for $pp \to h+\mathrm{jet}$ with $CP$-violating couplings}
In this appendix we collect the analytical expressions of the amplitudes contributing to $pp\to h +
\mathrm{jet}$ for $CP$-violating Higgs couplings. We consider first the one-loop amplitudes mediated
by the coupling $\tilde{\kappa}_{t}$ in Eq.~\eqref{lagrangian}.\par
The $gg\to hg$ amplitude can be
expressed in terms of helicity amplitudes \(\mathcal{M}_{\lambda_1\,\lambda_2\,\lambda_3}\), where
the \(\lambda_i\) denote the helicities of the incoming (\(i=1,2\)) and outgoing (\(i=3\))
gluons. Out of the eight possible helicity combinations only four are independent and related to the
remaining four amplitudes through
\(\mathcal{M}_{\lambda_1\,\lambda_2\,\lambda_3}=\mathcal{M}_{-\lambda_1\,-\lambda_2\,-\lambda_3}\).
The amplitudes are given by\footnote{We define $s=(p_{1}+p_{2})^{2}$, $t=(p_{1}-p_{3})^{2}$ and $u=(p_{1}-p_{4})^{2}$, where $p_{1,2}$ are the momenta of the ingoing gluons, $p_{3}$ is the momentum of the outgoing gluon and $p_{4}$ the momentum of the Higgs. Conservation of momentum is expressed as $p_{1}+p_{2}=p_{3}+p_{4}\,$.}
\begin{eqnarray}
  \label{eq:9}
  \mathcal{M}_{+\,+\,+}(s,t,u)&=& \mathcal{N}\,F_1(s,t,u)\,,\\
  \mathcal{M}_{+\,+\,-}(s,t,u)&=&\mathcal{N}\,F_1(s,u,t)\,,\\
  \mathcal{M}_{-\,+\,-}(s,t,u)&=&\mathcal{N}\,F_2(s,t,u)\,,\\
  \mathcal{M}_{-\,+\,+}(s,t,u)&=&\mathcal{N}\,F_3(s,t,u)\,,
\end{eqnarray}
where the form factors \(F_i\) are defined as
\begin{eqnarray}
  \label{eq:8}
  F_1(s,t,u)&=&\quad\sqrt{\frac{t}{s\,u}}\;\,\left[G(s,t)-G(s,u)+G(t,u)\right]\,,\\
  F_2(s,t,u)&=&-\frac{m_h^2}{\sqrt{s\,t\,u}}\left[G(s,t)+G(s,u)+G(t,u)\right]\,,\\
  F_3(s,t,u)&=&\;\;\;\sqrt{\frac{s}{t\,u}}\;\left[G(s,t)+G(s,u)-G(t,u)\right]\,,\\
  G(x,y)&=&x\, y\, D_0(x, y) + 2 x\,C_0(y) + 2 y\, C_0(x)\,,
\end{eqnarray}
with $C_{0}(x)\equiv C_{0}(0,x,m_{h}^{2},m_{t}^{2},m_{t}^{2},m_{t}^{2})$ and $D_{0}(x,y)\equiv
D_{0}(0,0,0,m_{h}^{2},x,y,m_{t}^{2},m_{t}^{2},m_{t}^{2},m_{t}^{2})$, where the scalar integrals are
given in the conventions of Ref.~\cite{Hahn:1998yk}. The common factor \(\mathcal{N}\) reads
\begin{equation}
  \label{eq:10}
  \mathcal{N}=\frac{\sqrt{3}\,\alpha_s^{3/2} m_t^2 \tilde{\kappa}_t}{\sqrt{\pi}\,v}\,.
\end{equation}
The unaveraged cross section is then given by
\begin{equation}
  \label{eq:12}
  \sum_{\tn{pol.}}\left|\mathcal{M}(gg\to hg)\right|^2=\sum_{\lambda_1\,\lambda_2\,\lambda_3} \left|\mathcal{M}_{\lambda_1\,\lambda_2\,\lambda_3}\right|^2.
\end{equation}
Similarly, the squared matrix element for \(q\bar{q}\to hg\) can be expressed in helicity amplitudes
\(\mathcal{M}^{q\bar{q}}_{\lambda_1\,\lambda_2\,\lambda_3}\) where \(\lambda_{1,2}\) now denote the
polarization of the incoming quark and anti-quark, respectively, and \(\lambda_3\) the helicity of the
outgoing gluon. The four non-zero combinations are related via
\begin{equation}
  \mathcal{M}^{q\bar{q}}_{R\,L\,+}(s,t,u)=-\mathcal{M}^{q\bar{q}}_{L\,R\,-}(s,t,u)=-{M}^{q\bar{q}}_{R\,L\,-}(s,u,t)=\mathcal{M}^{q\bar{q}}_{L\,R\,+}(s,u,t)
\end{equation}
and given by
\begin{equation}
  \mathcal{M}^{q\bar{q}}_{R\,L\,+}(s,t,u)=-\frac{2\sqrt{2}}{\sqrt{3\,s}}\,\mathcal{N}\, t\, C_0(s)\,.
\end{equation}
The unaveraged squared matrix element for the unpolarized cross section is then given by
\begin{equation}
  \label{eq:13}
  \sum_{\tn{pol.}}\left|\mathcal{M}_{q\bar{q}}(s,t,u)\right|^2=\sum_{\lambda_1\,\lambda_2\,\lambda_3}\left|\mathcal{M}^{q\bar{q}}_{\lambda_1\,\lambda_2\,\lambda_3}(s,t,u)\right|^2=\frac{16}{3}\mathcal{N}^{\,2}\frac{t^2+u^2}{s}\left|C_0(s)\right|^{2}.
\end{equation}
Finally, the matrix elements squared for the processes $qg\to qh$ and $\bar{q}g\to \bar{q}h$ are obtained via the permutations
\begin{equation}
\sum_{\tn{pol.}} \left| \mathcal{M}_{qg}(s,t,u)\right|^{2} = -\sum_{\tn{pol.}} \left| \mathcal{M}_{q\bar{q}}(u,t,s)\right|^{2}\,,\qquad \sum_{\tn{pol.}} \left| \mathcal{M}_{\bar{q}g}(s,t,u)\right|^{2} = -\sum_{\tn{pol.}} \left| \mathcal{M}_{q\bar{q}}(t,s,u)\right|^{2}\,, 
\end{equation}
respectively.\par
For a large mass of the fermion running in the loops, $m_{t}^{2}\gg s, -t, -u,
m_{h}^{2}\,$, we can expand the scalar functions $C_0(x)$ and $D_0(x,y)$ in  powers of
$1/m_{t}^{2}$. Keeping only the leading terms we have
\begin{equation}
  \label{eq:14}
  C_0(x) \to -\frac{1}{2\,m_t^2}\,,\qquad D_0(x,y)\to\frac{1}{6\,m_t^4}\,.
\end{equation}
In this limit the amplitudes $\mathcal{M}_{gg,q\bar{q}}$ simplify greatly: they are independent of $m_{t}\,$, and equal to the tree-level amplitudes computed using the effective coupling proportional to $\tilde{\kappa}_{g}$ in Eq.~\eqref{lagrangian} (the equality holds for $\tilde{\kappa}_{t}=\tilde{\kappa}_{g}$). The amplitudes squared and summed over polarizations take the simple form
\begin{equation}
\sum_{\tn{pol.}} \left| \mathcal{M}_{gg}\right|^{2} \,\to\, \frac{3g_{s}^{6}\tilde{\kappa}_{t}^{2}}{8\pi^{4}v^{2}}\,\frac{s^{4}+t^{4}+u^{4}+m_{h}^{8}}{stu}\,,\qquad \sum_{\tn{pol.}} \left| \mathcal{M}_{q\bar{q}}\right|^{2} \,\to\, \frac{g_{s}^{6}\tilde{\kappa}_{t}^{2}}{16\pi^{4}v^{2}}\,\frac{t^{2}+u^{2}}{s}\,.
\end{equation}

\end{document}